\documentclass[11pt]{article}
\usepackage{amsmath, amsfonts, amsthm, amssymb, amscd, mathrsfs}
\usepackage[mathcal]{euscript}
\usepackage{color}

\newcounter{mnotecount}[section]

\usepackage{fullpage}
\usepackage{bbm}
\usepackage{lastpage}
\usepackage{enumerate}
\usepackage{fancyhdr}
\usepackage{mathrsfs}
\usepackage{xcolor}
\usepackage{graphicx}
\usepackage{listings}
\usepackage{hyperref}
\usepackage{array}
\usepackage{tikz-cd}
\usepackage[utf8]{inputenc}
\usepackage[T1]{fontenc}
\usepackage{pdfpages}
\usepackage{titlesec}
\usepackage{etoolbox}
\usepackage{wrapfig}
\usepackage{caption}
\usepackage{subcaption}
\usepackage[toc,page]{appendix}
\usepackage{datetime}
\usepackage{verbatim}
\usepackage{mathtools}
\usepackage{tensor}
\usepackage{multicol}
\usepackage{enumitem}

\begin{document}

\theoremstyle{definition}
\newtheorem{exercise}{Exercise}[section]
\theoremstyle{plain}
\newtheorem{remark}{Remark}[section]
\theoremstyle{plain}
\newtheorem{prop}{Proposition}[section]
\theoremstyle{plain}
\newtheorem{thm}{Theorem}[section]
\theoremstyle{plain}
\newtheorem{lemma}{Lemma}[section]
\theoremstyle{definition}
\newtheorem{defn/}{Definition}[section]

\newenvironment{defn}
{\renewcommand{\qedsymbol}{$\diamond$}%
	\pushQED{\qed}\begin{defn/}}
	{\popQED\end{defn/}}

\theoremstyle{plain}
\newtheorem{cor}{Corollary}[section]
\newtheorem{example/}{Example}[section]

\newenvironment{example}
{\renewcommand{\qedsymbol}{$\diamond$}%
	\pushQED{\qed}\begin{example/}}
	{\popQED\end{example/}}

\newcommand{\N}{\mathbb{N}}
\newcommand{\C}{\mathbb{C}}
\newcommand{\R}{\mathbb{R}}
\newcommand{\Z}{\mathbb{Z}}
\newcommand{\Q}{\mathbb{Q}}
\newcommand{\Primes}{\mathbb{P}}
\newcommand{\csigma}{\mathfrak{S}}
\newcommand{\1}{\mathbbm{1}}
\newcommand{\bb}[1]{\mathbb{#1}}
\newcommand{\mcal}[1]{\mathcal{#1}}
\newcommand{\T}{\mathbb{T}}
\newcommand{\A}{\mathbb{A}}
\newcommand\sbullet[1][.5]{\mathbin{\vcenter{\hbox{\scalebox{#1}{$\bullet$}}}}}


\title{Improved decay estimates and $C^2$-asymptotic stability of solutions to the Einstein-scalar field system in spherical symmetry} 

\author{Jo\~ao L. Costa$^{(2,1)}$, Rodrigo L. Duarte$^{(1)}$ and Filipe C. Mena$^{(1)}$\\\\
{\small $^{(1)}$Centro de An\'alise Matem\'atica, Geometria e Sistemas Din\^amicos,}
\\
{\small Instituto Superior T\'ecnico, Universidade de Lisboa, Av. Rovisco Pais 1, 1049-001 Lisboa, Portugal}\\
{\small $^{(2)}$Instituto Universit\'ario de Lisboa (ISCTE-IUL), Av. das For\c{c}as Armadas, 1649-026 Lisboa, Portugal}\\
}

\maketitle

\begin{abstract}
We investigate the asymptotic stability of solutions to the characteristic initial value problem for the Einstein (massless) scalar field system with a positive cosmological constant. We prescribe spherically symmetric initial data on a future null cone with a wider range of decaying profiles than previously considered. New estimates are then derived in order to prove that, for small data, the system has a unique global classical solution. We also show that the solution decays exponentially in (Bondi) time and that the radial decay is essentially polynomial, although containing logarithmic factors in some special cases. This improved asymptotic analysis allows us to show that, under appropriate and natural decaying conditions on the initial data,  the future asymptotic solution is differentiable, up to and including spatial null-infinity,  and approaches the  de Sitter solution, uniformly, in a neighborhood of infinity. Moreover, we analyze the decay of derivatives of the solution up to second order showing the (uniform) $C^2$-asymptotic stability of the de Sitter attractor in this setting. This corresponds to a surprisingly strong realization of the cosmic no-hair conjecture.
\end{abstract}
\newpage
\tableofcontents
\newpage
\section{Introduction}

\subsection{Motivation and previous results}

The Einstein field equations are non-linear second order partial differential equations for a Lorentzian metric of a 4-dimensional manifold called spacetime. They constitute the basis of General Relativity theory and are some of the most fascinating and  challenging differential equations that originate from physics. A  main difficulty in analyzing these equations is that they are defined on a manifold while, at the same time, they determine the metric of the manifold itself as they evolve. 

An important conjecture in the theory of General Relativity is the so-called cosmic no-hair conjecture which roughly says that generic expanding solutions to the Einstein equations with a positive cosmological constant $\Lambda$ asymptote to the de Sitter solution -- the maximally symmetric vacuum solution of Einstein's field equations with a positive cosmological constant. In turn, the inclusion of $\Lambda>0$ is motivated by current observations in cosmology (see~\cite{ringOx} for a mathematically inclined introduction to the subject).

This conjecture is usually attributed to Gibbons and Hawking~\cite{Gibbons-Hawking} and its earlier history was mostly concerned with the study of homogenous  and isotropic cosmologies, whose dynamical content is encoded on ODEs; a notable example of this period are the celebrated results of Wald~\cite{Wald}.
In the meantime, far reaching results, concerning more general dynamics,  have been produced, with considerable increase of activity in recent years. Most of these results concern the analysis of geometric PDEs in the context of the study of perturbations of spatially homogeneous cosmological models, see e.g. \cite{Alho, ringCNH, Costa-Natario-Oliveira, fajman,Forno, Friedrich, Fried-dust, gasperin, Hadzic, Valiente, Nungesser,Rendall, ringOx, Rod-Speck,Speck} and references therein.

The cosmic no-hair conjecture can be rephrased more precisely as a statement about the existence of a global attractor, the de Sitter solution, for a generic class of solutions to the Einstein equations with a positive cosmological constant, coupled to appropriate matter fields. Global attractor here meaning that, all solutions, in a generic class, become close to the de Sitter solution, near  (null) infinity. A stronger version of the conjecture emanates from the expectation that, in some cases, the attractor is moreover {\em asymptotically stable}, meaning that all solutions, in the generic class, converge to de Sitter, in some appropriate norm, as we approach infinity~\footnote{It is well known that the Nariai solution provides a simple counterexample for these claims; but since this solution is known to be unstable~\cite{fajman}, cosmic no-hair is expected to hold generically. }.  

Most of the previously referred results, directly or indirectly, provide a realization of the first (weaker) version of the conjecture. Notably, the recent results of Fournodavlos~\cite{Forno}, also concerning the Einstein-scalar field system, but without symmetry assumptions, provides a very clear and detailed description of the asymptotic behavior of solutions: in particular, they show that the spacetime metric of such solutions approaches, along a specific spacelike foliation, a metric that, near  infinity, can be made arbitrarily close to de Sitter metric, if we choose the initial data to also be arbitrarily close to de Sitter~\footnote{We refer to this excellent paper~\cite{Forno} for clarifications concerning all missing details.}.  

The {\em asymptotically stable} version of the conjecture has also seen some relevant developments recently, propelled by a reformulation of the conjecture proposed, and very clearly formulated, in~\cite{ringCNH}. The main insight of this new formulation can be seen as a weakening of the concept of asymptotic stability, where one ``only'' demands that the approach to de Sitter happens along the causal past of every observer (timelike curve) that reaches infinity. Although not as strong as a version of cosmic no-hair where the approach to de Sitter happens uniformly, in a neighborhood of infinity, this might correspond to a necessary compromise to truly capture, in full generality, the generic asymptotic behavior of solutions to the Einstein equations with a positive cosmological constant~\footnote{An heuristic argument corroborating this view, based on solutions to the linear wave equation and the phenomena of ``cosmic silence'', was proposed in the introduction to~\cite{Costa-Natario-Oliveira}. Moreover, in this paper, a version of the conjecture at hand was established in the context of spherically symmetric de Sitter black hole spacetimes.}.   

In contrast to this latter trend, in this paper, we will present results where the ($C^2$) asymptotic stability of de Sitter is established uniformly in a  neighborhood of infinity. This corresponds to a particularly strong realization of the cosmic no-hair conjecture. Nonetheless, our results are restricted to the spherically symmetric setting and explore the possibilities created by the Bondi gauge which, in particular, provides a natural foliation of  a neighborhood of null infinity in terms of surfaces of constant area radius. If our results are specific to spherical symmetry remains to be seen~\footnote{The fact that a natural choice of foliation, like the one provided by the Bondi gauge in spherical symmetry, does not seem to exist for general solutions creates an obstacle to extend our results to the non-symmetrical setting. See~\cite{schlue2} for a discussion of this particular issue in a related context.}. 

Among the physically interesting source fields that we can couple to the Einstein equations are scalar fields, which can be motivated by particle physics \cite{Arbey}, but also by the fact that, in our spherically symmetric setting, they provide extra dynamical degrees of freedom that mimic those of vacuum in the non-symmetrical case. 

Following the pioneering work of Christodoulou \cite{Christodoulou:1986} on the analysis of the spherically symmetric massless  Einstein-scalar field system in Bondi coordinates, a series of  papers has started in \cite{CostaProblem} with the goal of analysing the system with a positive cosmological constant \cite{CostaReview,Costa-Natario-Oliveira,Costa-Mena}. However, regarding asymptotic stability, the end result of this series was restricted to a $C^0$ analysis, which is manifestly unsatisfactory. In fact, since the spacetime curvature involves second order derivatives of the metric, a $C^2$ analysis is important for a full geometric characterization of the spacetime.  This is achieved in the current paper.

As was already mentioned, the global existence and uniqueness, and the $C^0-$asymptotic stability of solutions to the Einstein-scalar field system was proved  in \cite{Costa-Mena} for small, spherically symmetric initial data, with radial decay of $O(r^{2-\delta})$, $\delta>0$, along a characteristic initial surface.  The main goal of the current paper is to upgrade these results by showing the (uniform) $C^2-$asymptotic stability of solutions. It turns out that our techniques to control some second order derivatives seem to require initial data with faster decay, that is $O(r^{2-\delta})$, with $\delta<0$. Therefore, a considerable amount of new work (see Section~\ref{challenges}) has to be done in order to revisit the global well-posedness results of~\cite{Costa-Mena} and more so to control the asymptotic behavior of solutions and their derivatives up to second order. By enlarging the set of initial data and doing a $C^2-$asymptotic analysis we are also testing the robustness of the methods developed in~\cite{Costa-Mena} to establish the $C^0$ stability. 

We will describe in section  \ref{challenges} the main technical challenges that emerge in this problem but, before that, let us  present the integro-differential system that we will be using throughout this paper. 
\subsection{The Einstein-scalar field integro-differential system}
Consider a Lorentzian 4-dimensional manifold \((M,g)\) and a massless scalar field \(\phi\) defined on $M$. The Einstein-scalar field system with a cosmological constant $\Lambda>0$ is given by
\begin{equation}\label{eq1}
\begin{dcases}
R_{\mu \nu}-\frac{1}{2}Rg_{\mu \nu}+\Lambda g_{\mu \nu}=\partial_\mu \phi \partial_\nu \phi-\frac{1}{2}g_{\mu \nu}\tensor{g}{^\gamma^\rho}\partial_\gamma \phi \partial_\rho \phi;\\
\square_g \phi=0.
\end{dcases}
\end{equation}
where $R$ is the Ricci scalar curvature, $R_{\mu\nu}$ denotes the components of the Ricci tensor and $g_{\mu\nu}$ the components of metric $g$. 
Without loss of generality we will assume that \(\Lambda=3\). We consider spherically symmetric spacetimes $(M,g)$ with a metric in the so-called Bondi coordinates as
\begin{equation}
\label{metric}
g=-f(u,r)\tilde{f}(u,r)du^2-2f(u,r)dudr+r^2\sigma_{\mathbb{S}^2},
\end{equation}
where \(\sigma_{S^2}=d\theta^2+\sin^2(\theta)d\varphi^2\) is the metric on the 2-sphere and the coordinate ranges are \((u,r)\in [0,+\infty[\times [0,+\infty[\). Then the system \eqref{eq1} becomes
\begin{equation}\label{eq2}
\begin{dcases}
\frac{1}{f}\partial_rf=\frac{r}{2}(\partial_r\phi)^2\\
\partial_r(r\tilde{f})=(1-3r^2)f\\
\frac{1}{r}\left(\partial_u-\frac{\tilde{f}}{2}\partial_r\right)\partial_r(r\phi)=\frac{1}{2}\partial_r\tilde{f}\partial_r\phi.
\end{dcases}
\end{equation}
An important particular case of metric \eqref{metric} is the de Sitter metric $g^{\mathrm {dS}}$ given by the choice $f(u,r)=f^{\mathrm {dS}}(u,r):=1$ and $\tilde f(u,r)=\tilde f^{\mathrm {dS}}(u,r):=1-\Lambda r^{2}/3$, i.e. 
\begin{equation}
g^{\mathrm {dS}}=-\left(1-r^{2}\right)du^{2}-2dudr+r^{2}\sigma_{\mathbb{S}^2}.
\label{dSBondi}
\end{equation}
We recall that following \cite{Christodoulou:1986} and setting 
$$h:=\partial_r(r\phi),$$
the system \eqref{eq2} can be reduced to the integro-differential equation
\begin{equation}\label{eq3}
Dh=G(h-\bar{h}),
\end{equation}
with the definitions
\begin{equation}
\aligned
D:=&\partial_u-\frac{1}{2}\tilde{f}\partial_r,\\
G:= &\frac{1}{2}\partial_r\tilde{f}\\
\bar{h}(u,r):=& \frac{1}{r}\int_0^rh(u,s)ds.
\endaligned
\end{equation}
We will assume, without loss of generality, that \(f(u,0)=1\) which corresponds to a rescaling of the time coordinate $u$. 
In the following lemma we summarize some basic properties of the above quantities. 
\begin{lemma}\label{lemma1}
If system \eqref{eq2} is satisfied, then the following hold:
\begin{align}
&\bar{h}(u,r)=\phi(u,r)\label{eq4};\\
&\partial_r\bar{h}=\frac{h-\bar{h}}{r}\label{eq5};\\
&f(u,r)=\exp\left(\frac{1}{2}\int_0^r\frac{(h(u,s)-\bar{h}(u,s))^2}{s}ds\right)\label{eq6};\\
&\tilde{f}(u,r)=\frac{1}{r}\int_0^r(1-3s^2)f(u,s)ds=\bar{f}(u,r)-\frac{3}{r}\int_0^rs^2f(u,s)ds; \label{eq7}\\
&G=\frac{1}{2r}(f-\tilde{f}-3r^2f)=\frac{1}{2r}(f-\bar{f}+\frac{3}{r}\int_0^rs^2f(u,s)ds-3fr^2). \label{eq8}
\end{align}	
\end{lemma}
It is known \cite{CostaProblem, Christodoulou:1986} that if \((g,\phi)\) satisfy the system \eqref{eq1}, then \(h\) defined by \(h=\partial_r(r\phi)\) satisfies equation \eqref{eq3}. Conversely, if \(h\) solves \eqref{eq3} then \(\phi, f, \tilde{f}\) satisfy \eqref{eq1} and are related to \(h\) by \eqref{eq4}, \eqref{eq6}, \eqref{eq7}, respectively.

If we now differentiate equation \eqref{eq3} with respect to \(r\) we obtain
\[
\partial_u\partial_rh-\frac{1}{2}\partial_r\tilde{f}\partial_rh-\frac{1}{2}\tilde{f}\partial_r^2h=\frac{1}{2}\partial_r^2\tilde{f}(h-\bar{h})+\frac{1}{2}\partial_r \tilde{f}(\partial_rh-\partial_r\bar{h}).
\]
Rearranging and using \eqref{eq5} we obtain 
\begin{equation}\label{eq46}
(D-2G)(\partial_rh)=-J(\partial_r\bar{h}),
\end{equation}
where
\[
J:=-\frac{r}{2}\partial_r^2\tilde{f}+\frac{1}{2}\partial_r\tilde{f}
\]
By using the above definitions we arrive at the useful expression:
\begin{equation}
\label{diff-J}
J= 
G-r\partial_rG
= 3G+3rf-\frac{1-3r^2}{2}\partial_rf.
\end{equation}
which will be used later to estimate $J$.
\subsection{Main results}
Given $\delta\in \R$ we define
\begin{equation}
\label{jumps}
\delta^+=\begin{dcases}
\delta,&\text{ if }\delta>0\\
0,&\text{ if }\delta\leq 0,
\end{dcases} 
\quad \quad \chi(\delta)=
\begin{dcases}
1,&\text{ if }\delta=0\\
0,&\text{ if }\delta\neq 0,
\end{dcases}
\quad \quad H(\delta)=
\begin{dcases}
1,&\text{ if }\delta>0\\
0,&\text{ if }\delta\le 0\; 
\end{dcases}\;.
\end{equation}
The following three theorems constitute the main results of this paper:
\begin{thm}[Global existence]
\label{thm-existence}
Let \(-1< \delta<1/2\). Suppose that \(\phi_0\in C^{k+2}([0,+\infty[)\) satisfies
\[
\sup_{r\geq 0}\left(|\partial_r(r\phi_0)(r)|+|(1+r)^{2-\delta}\partial_r^2(r\phi_0)(r)|+|(1+r)^{3-\delta}\partial_r^3(r\phi_0)(r)|\right)<\infty.
\]
There exists some \(\varepsilon_0>0\) such that, when
\begin{equation}
\label{smallData}
\sup_{r\geq 0}|(1+r)^{2-\delta}\partial_r^2(r\phi_0)(r)|<\varepsilon_0,
\end{equation}
then there exists a unique solution \((M=\R_0^+\times \R_0^+\times \mathbb{S}^2, g, \phi)\) to the Einstein-scalar field system \eqref{eq1}. Here \((M,g)\) is a \(C^{k+1}\) spherically symmetric spacetime in Bondi coordinates and \(\phi\in C^{k+1}(M)\) satisfies the initial condition
\[
\phi(0,r)=\phi_0(r),\ \forall r\geq 0.
\]
\end{thm}
\begin{thm} [Estimates for the matter field]
\label{thm-decay}
Under the conditions of Theorem \ref{thm-existence}, if we set Bondi time to be the proper time of the observer at the center of symmetry, i.e. \(f(u, r=0)\equiv 1\), then there exists a continuous function \(\underline{\phi}:[0,+\infty[\rightarrow \R\) which is differentiable in $]0,+\infty[$ and $\gamma=\gamma(\delta)>0$ such that
\begin{align}
|\phi(u,r,\omega)-\underline{\phi}(u)|&\lesssim \frac{\log^{\chi(\delta)}(e+r)}{(1+r)^{1-\delta^+}}e^{-(1+\delta/2)u} \label{eq50}\\
|\partial_r\phi(u,r)|&\lesssim \frac{\log^{\chi(\delta)}(e+r)}{(1+r)^{2-\delta^+}}e^{-(1+\delta/2)u} \label{eq51}\\
 |\partial_u\phi(u,r)| & \lesssim  (1+r)^{\delta^+} \log ^{\chi(\delta)} (e+r) e^{-(1+\delta/2)u} \label{eq51a}\\
|\partial_r^2\phi(u,r)|&\lesssim \frac{\log^{\chi(\delta)}(e+r)}{(1+r)^{3-\delta^+}}e^{-(1+\delta/2)u} \label{eq52}\\
 |\partial^2_u\phi(u,r)|& \lesssim (1+r)^{1+\delta} e^{-\gamma u} [\log{(e+r)}]^{4\chi(\delta)+1-H(\delta)} \label{eq52a}\;.
\end{align}
Also, there exists \(\underline{\phi}(\infty)\in \R\) such that, given \(R>0\), if \(\varepsilon_0\leq \underline{\varepsilon}(R)\), with the latter sufficiently small, then there exists a constant \(C_R>0\) such that 
\begin{equation}\label{eq53}
\sup_{r\leq R}(|\phi(u,r)-\underline{\phi}(\infty)|+|\partial_r\phi(u,r)|+|\partial_r^2\phi(u,r)|)\leq C_Re^{-2u}.
\end{equation}
\end{thm}
\begin{remark}
Note that the estimates for $|\partial_u\phi(u,r)|$ and $ |\partial^2_u\phi(u,r)|$ do not show decay in $r$. This is due to the fact that the vector field $\partial_u$ has a diverging norm, when $r\rightarrow+\infty$, in fact $g_{uu}\sim r^2$ in that asymptotic regime. 
\end{remark}
\begin{thm} [$C^{2}$ -- asymptotic stability]
\label{thm-stability}
Assume the conditions of Theorem \ref{thm-existence}.
Let \(R\gg 1\). Fix Bondi time by imposing \(d\hat{u}=f(u,r=\infty)du\). Then, there exists an orthonormal frame \((e_I)_{I=0,1,2,3}\), in the \(\{r>R\}\) region of de Sitter spacetime, and  a diffeomorphism mapping this region to the region \(\{r>R\}\) in our spacetime such that, by writing \(g_{IJ}(\hat{u},r)=g(e_I,e_J)\) and $g_{IJ}^{dS}:=g^{dS}(e_I,e_J)=\eta_{IJ}$,~\footnote{Recall the usual convention $\eta_{00}=-1$, $\eta_{ii}=1$, for $i=1,2,3$ and $\eta_{IJ}=0$, if $I\neq J$.} we have 
\begin{equation}\label{eq54}
|g_{IJ}(\hat{u},r)-g_{IJ}^{dS}|\lesssim \frac{\log^{2\chi(\delta)}(e+r)}{(1+r)^{2(1-\delta^+)}}e^{-2(1-\varepsilon)(1+\delta/2)\hat{u}}\;,
\end{equation}
where $\varepsilon>0$ can be made arbitrarily small by decreasing the size of the initial data.  Consequently \((M,g)\) is geodesically complete towards the future. 

Furthermore, we have 

\begin{equation}
\label{estimates}
\aligned
 |\partial_{e_0}g_{IJ}|&\lesssim \frac{[\log(e+r)]^{7\chi(\delta)+1-H(\delta)}}{(1+r)^{2-2\delta^+}}e^{-(1+\delta/2)(1-\varepsilon)\hat{u}}\\
   |\partial_{e_1}g_{IJ}|&\lesssim \frac{[\log(e+r)]^{7\chi(\delta)+1-H(\delta)}}{(1+r)^{2-\delta^+-\delta}}e^{-(1+\delta/2)(1-\varepsilon)\hat{u}}
\endaligned
\end{equation}
and, for \(-1<\delta<0\), there exists $\gamma=\gamma(\delta,\varepsilon)>0$ for which
\begin{equation}
\label{estimates2}
\aligned   
    |\partial^2_{e_1}g_{IJ}|, ~|\partial_{e_0}\partial_{e_1}g_{IJ}| &\lesssim \frac{\log^2(e+r)}{(1+r)^{2-\delta}}e^{-\gamma\hat{u}},
    \\
    |\partial^2_{e_0}g_{IJ}|&\lesssim \frac{\log^2(e+r)}{(1+r)^2}e^{-\gamma\hat{u}}\;.
\endaligned
\end{equation}
\end{thm}
\begin{remark}
The previous theorem provides a surprisingly strong realization of the cosmic no-hair conjecture by establishing that the spacetime metric of our solutions converges, {\bf uniformly} in a neighborhood of null infinity ($r=+\infty$), to the de Sitter metric, in a $C^2$ norm. 
\end{remark}
The proof of Theorem \ref{thm-existence}, Theorem \ref{thm-decay} and Theorem \ref{thm-stability} can be found in Section \ref{global-sec}, Section \ref{decay-sec} and Section \ref{converge-sec}, respectively.

By inspection of the presented results we see that, most of them, improve when we have a faster initial radial decay rate, corresponding to $\delta<0$. These improvements include, higher regularity of the scalar field at null infinity, asymptotic stability in a stronger $C^2$ norm and faster spacetime radial decay rates, for all relevant quantities.  
Nonetheless,  using a simple domain of dependence argument, we can extend all these properties to an arbitrarily large domain of all (global) solutions, even those starting from slower decaying initial data: To do that, assume we are dealing with a solution emanating from data with decay parametrized by $\delta\in(-1,1/2)$ and for every $R>0$, denote the initial null cone truncated at $r=R$ by ${\cal C}_R:=\{u=0\}\cap\{r<R\}$. Now, let $R_1>R>R_0$, with $R_0$ sufficiently large and depending on the value of $\varepsilon_0$ in~\eqref{smallData},  and modify the initial data by imposing a faster decaying rate of $O(r^{2-\delta_1})$, with $\delta_1<0$, while preserving the original data in  ${\cal C}_{R_1}$. In view of our estimates for the ingoing light rays, established in Section~\ref{sectionChar}, we see that, by enlarging $R_1-R$ if necessary, we have $D_{\delta}^+({\cal C}_R)\subset D_{\delta_1}^+({\cal C}_{R_1})$, where $D^+_{\delta}$ represents the future domain of dependence, seen as a subset of the $(u,r)$ plane, with respect to the metric of the original solution (with decay parametrized by $\delta$) and  $D^+_{\delta_1}$ is the future domain of dependence of the modified solution (with decay parametrized by $\delta_1$). By uniqueness, both solutions coincide in $D_{\delta}^+({\cal C}_R)$ and the desired result holds. For future reference we collect these conclusions in the following: 
\begin{cor}
\label{corIntro}
Under the conditions of Theorem~\ref{thm-existence},  there exists $R_0$ depending on the value of $\varepsilon_0$ in~\eqref{smallData}, such that, for all $R>R_0$ and all $\delta_1\in(-1,0)$, all the conclusions of Theorem~\ref{thm-decay} and Theorem~\ref{thm-stability} hold in $D_{\delta}^+({\cal C}_R)$, with  $\delta$ replaced by $\delta_1$.
\end{cor}

\begin{remark}
The last discussion may raise the question of what is the ``correct'' initial radial decay rate. The answer depends on the  problem under consideration. If one is interested in solving a Cauchy problem with (regular) initial data posed on a hypersurface with $\mathbb S^3$ topology, then our results suggest that this should evolve into characteristic data, posed in $u=0$, with radial decay rate with $\delta$ arbitrarily close to $-1$. But if one is interested in an open universe  model with $\mathbb R^3$ topology, then the initial radial decay  can be prescribed freely and, in that spirit, it is interesting to study solutions with a wider range of initial decays; that is, in essence, what we do.    
\end{remark}
%
\subsection{Challenges and outline of the paper}
\label{challenges}
The $C^2$ stability analysis requires substantial new work with respect to the previous results in~\cite{Costa-Mena}. In particular we note that:
 \begin{enumerate}
 \item We require more general initial data than \cite{Costa-Mena} with a wider range of radial decay. In particular, $\delta$ is now allowed to be zero or negative, which corresponds to a faster decay. Note also that the $\delta=0$ case introduces logarithmic factors in the estimates of most relevant quantities; a nuisance that one has to deal with, in this particular case.  
\item We need new estimates for some crucial quantities, for example for $\partial_r J$, where $J$ is defined in \eqref{diff-J}.
 This is now necessary, for example, in the existence proof of Theorem \ref{thm-existence}. In fact, in order to close the contraction argument, one needs to propagate the faster decay rate of the initial data (in our setup, this corresponds to the $\delta<0$ case) to the entire spacetime, which, in turn, requires a more careful control of several key quantities.  
 \item We need to derive new estimates for the derivatives of the metric and related quantities up to second order. As already discussed this translates into stronger asymptotic stability results for the future attractor solution (the de Sitter solution).
\item We prove new differentiability properties of the asymptotic solution, up to and including $r=+\infty$, which corresponds to future null-infinity. However our differentiability proof requires stronger initial decay rates than what was considered in \cite{Costa-Mena}. 
Nonetheless,  as already discussed, using a simple domain of dependence argument, we can extend this property to all (global) solutions, even those starting from slower decaying initial data.
 \end{enumerate}

The outline of the paper is as follows: In Section \ref{bounds} we give preliminary estimates of some crucial quantities as well as new estimates along the characteristics involving the new logarithmic terms. In Section \ref{global-sec} we revisit the global existence results of \cite{Costa-Mena} by including the more general decay.  In Section \ref{decay-sec} we prove new global decay estimates for the matter and geometric quantities and establish important asymptotic properties of the solutions such as its differentiability. Finally, in Section \ref{converge-sec} we show the asymptotic convergence of the solutions to the de Sitter spacetime as well as their $C^2-$asymptotic stability.
\section{A Priori Bounds}
\label{bounds}
\subsection{Basic estimates and norms}
Consider $U\in~ ]0,+\infty], \delta \in \R$ and $w:[0,U[\times[0,+\infty[\to \R$ a continuous function.
In what follows we will make use of the norms
\[
\|w(u,\cdot)\|_{L_r^{\infty,2-\delta}}:=\sup_{r\geq 0}|(1+r)^{2-\delta}w(u,r)|
\]
and
\[
\|w\|_{L_U^\infty L_r^{\infty,2-\delta}}:=\sup_{0\leq u\leq U}\|w(u,\cdot)\|_{L_r^{\infty,2-\delta}}.
\]
As a matter of notation we will often write \(L_r^\infty\) instead of \(L_r^{\infty,0}\). 
\begin{lemma}\label{lemma2.1}
Let \(\delta\in ]-\infty, 1[\). For a sufficiently regular function \(w\) we have
\[
|w(u,r)-\bar{w}(u,r)|\leq C\frac{r\log^{\chi(\delta)}(e+r)}{(1+r)^{2-\delta^+}}\|\partial_rw(u,\cdot)\|_{L_r^{\infty, 2-\delta}},
\]
for some constant \(C=C(\delta)>0\).
\end{lemma}
\begin{proof}
In general, for \(p\in \R\), we have
\[
\begin{split}
|w(u,r)-\bar{w}(u,r)|&=\frac{1}{r}\left|\int_0^r[w(u,r)-w(u,s)]ds\right|=\frac{1}{r}\left|\int_0^r\int_s^r\partial_rw(u,\rho)d\rho ds \right|\\
&\leq \frac{1}{r}\int_0^r\int_s^r\frac{1}{(1+\rho)^p}d\rho ds \|\partial_rw(u,\cdot)\|_{L_r^{\infty,p}}\;.
\end{split}
\]
In particular, when \(p=0\) we get
\begin{align}
|w(u,r)-\bar{w}(u,r)|&\leq \frac{r}{2}\|\partial_rw(u,\cdot)\|_{L_r^{\infty}}\label{eq9}
\end{align}
One can explicitly compute
\[
\frac{1}{r}\int_0^r\int_s^r\frac{1}{(1+\rho)^2}d\rho ds=\frac{\log(1+r)}{r}-\frac{1}{1+r}
\]
to obtain
\begin{equation}\label{eq11}
|w(u,r)-\bar{w}(u,r)|\leq A(r)\frac{r\log(e+r)}{(1+r)^2}\|\partial_r w(u,\cdot)\|_{L_r^{\infty,2}},
\end{equation}
where
\[
A(r)=\frac{(1+r)^2}{r\log(e+r)}\left[\frac{\log(1+r)}{r}-\frac{1}{1+r}\right].
\]
One can check that \(\lim_{r\rightarrow \infty}A(r)=1\), so there is some \(r_0\) such that
\[
r\geq r_0\implies 0\leq A(r)\leq \frac{3}{2}\;.
\]
In particular from \eqref{eq11}
\[
\aligned
r\geq r_0\implies |w(u,r)-\bar{w}(u,r)|
\leq & \frac{3}{2}\frac{r\log(e+r)}{(1+r)^2}\|\partial_rw(u,\cdot)\|_{L_r^{\infty,2}}\;.
\endaligned
\]
On the other hand, when \(0\leq r\leq r_0\), by continuity there is some \(C'\) such that
\[
\frac{(1+r)^2}{2\log(e+r)}\leq C', \forall r\in [0,r_0].
\]
Also, note that \(\|\partial_rw(u,\cdot)\|_{L_r^\infty}\leq \|\partial_rw(u,\cdot)\|_{L_r^{\infty, 2}}\). Therefore, when \(0\leq r\leq r_0\) we have, using \eqref{eq9},
\[
|w(u,r)-\bar{w}(u,r)|\leq \frac{r}{2}\|\partial_rw(u,\cdot)\|_{L_r^{\infty}}\leq C'\frac{r\log(e+r)}{(1+r)^2}\|\partial_rw(u,\cdot)\|_{L_r^{\infty,2}}\;.
\]
The result now follows with \(C=\max\{C', 3/2\}\). The proof for the cases \(0<\delta<1\) and \(\delta<0\) is analogous.
\end{proof}
%
\subsection{Estimates for the crucial quantities $G$ and $J$}
Using Lemma \ref{lemma2.1} we have
\[
(h(u,r)-\bar{h}(u,r))^2\leq C\frac{r^2\log^{2\chi(\delta)}(e + r)}{(1+r)^{4-2\delta^+}}\|\partial_rh\|_{L_U^\infty L_r^{\infty,2-\delta}}^2.
\]
Throughout this paper we use the convention that the constant \(C\) may change from step to step. To ease the notation we set 
$$
X_\delta=\|\partial_rh\|_{L_U^\infty L_r^{\infty,2-\delta}}.
$$ 
From \eqref{eq6}, we get
\begin{equation}\label{eq16}
1\leq f(u,r)= \exp\left(\frac{1}{2}\int_0^r\frac{(h(u,s)-\bar{h}(u,s))^2}{s}ds\right)\leq e^{CX_\delta^2}\leq 1+\varepsilon_{X_\delta},
\end{equation}
where \(\varepsilon_{X_\delta}\) denotes a positive expression such that \(\varepsilon_{X_\delta}\rightarrow 0\) as\footnote{The use of the notation \(\varepsilon_{X_\delta}\) is similar to the usage of the little-o notation, in that it may represent different expressions from step to step and even in different appearances in the same step.} \(X_\delta\rightarrow 0\). Hence there is some \(\gamma>0\) such that, if \(X_\delta<\gamma\), then \(f\leq 2\). Also from \eqref{eq6}, we have \(0\leq \partial_rf=f(h-\bar{h})^2/(2r)\), so we obtain
\begin{equation}\label{eq12}
0\leq \partial_rf\leq C\frac{r\log^{2\chi(\delta)}(e+r)}{(1+r)^{4-2\delta^+}}X_\delta^2.
\end{equation}
\begin{prop}
\label{prop1}
Let \(\delta\in[-1,1/2[\). Then, there is some \(\gamma>0\) such that, if \(X_\delta<\gamma\), the following estimates hold:~\footnote{Here the notation \(X\lesssim Y\) means \(X\leq CY\), where $C$ is a positive constant.}
\begin{align}
&G(u,r)\leq -(1-\varepsilon_{X_\delta})r\label{eq13}\\
&|G(u,r)|\leq (1+\varepsilon_{X_\delta})r\label{eq14}\\
&|J(u,r)|\lesssim \frac{[\log(e+r)]^{2\chi(\delta)+1-H(\delta)}}{(1+r)^{1-2\delta^+}}X_\delta^2\label{eq15}\\
&|\partial_rJ(u,r)|\lesssim \frac{[\log(e+r)]^{2\chi(\delta)+1-H(\delta)}}{(1+r)^{2-2\delta^+}}X_\delta^2,\text{ for }r\geq 1\label{eq24}\;,
\end{align}
where $\delta^+$, $\chi(\delta)$ and $H(\delta)$ are defined in \eqref{jumps}.
\end{prop}
\begin{proof}
The proof of \eqref{eq13} and \eqref{eq14} mimics the one in Lemma 1 of \cite{Costa-Mena} by using \eqref{eq8} and \eqref{eq16} and noting that, since \(\partial_rf\geq 0\), we have
\[
\begin{split}
0\leq f-\bar{f}&=\frac{1}{r}\int_0^r\int_s^r\partial_rf(u,\rho)d\rho ds\leq C\frac{1}{r}\int_0^r\int_s^r\frac{\rho\log^{2\chi(\delta)}(e+\rho)}{(1+\rho)^{4-2\delta^+}}X_\delta^2d\rho ds\\
&\leq C\frac{1}{r}X_\delta^2\int_0^r\int_s^r\rho d\rho ds\leq CX_\delta^2r^2.
\end{split}
\]
So now we turn to \eqref{eq15}. Since \(f\) is increasing in \(r\),  we get
\[
|F(r)|:=\left|\frac{3}{r^2}\int_0^rs^2(f(u,s)-f(u,r))ds\right|\leq \frac{3}{r^2}\int_0^rs^2(f(u,r)-f(u,s))ds.
\]
For \(s<r\) there is some \(s<\rho<r\) such that $f(u,r)-f(u,s)=\partial_rf(u,\rho)(r-s)$. So, we have the estimate
\[
\begin{split}
f(u,r)-f(u,s)&\leq C\frac{\rho \log^{2\chi(\delta)}(e+\rho)}{(1+\rho)^{4-2\delta^+}}X_\delta^2r\leq C\frac{r\log^{2\chi(\delta)}(e+r)}{(1+s)^{3-2\delta^+}}X_\delta^2.
\end{split}
\]
We can now use this to get an estimate for \(F\) as
\[
\begin{split}
|F(r)|&\leq \frac{3}{r^2}\int_0^rs^2C\frac{r\log^{2\chi(\delta)}(e+r)}{(1+s)^{3-2\delta^+}}X_\delta^2ds=CX_\delta^2\frac{\log^{2\chi(\delta)}(e+r)}{r}\int_0^r\frac{s^2}{(1+s)^{3-2\delta^+}}ds\\
&\leq CX_\delta^2\frac{\log^{2\chi(\delta)}(e+r)}{r}\left[(1-H(\delta))\log(1+r)+\frac{H(\delta)}{2(1-\delta^+)(1-2\delta^+)\delta^+}\left((1+r)^{2\delta^+}-1\right)\right.\\
&\quad\left.-\frac{1}{2-2\delta^+}\frac{r^2}{(1+r)^{2-2\delta^+}}+\frac{1}{(1-\delta^+)(2\delta^+-1)}\frac{r}{(1+r)^{1-2\delta^+}}\right]\\
&\leq CX_\delta^2\frac{(1+r)^{2\delta^+}[\log(e+r)]^{2\chi(\delta)+1-H(\delta)}}{r}.
\end{split}
\]
For \(r\geq 1\) we have $1/r \leq 2/(1+r)$, 
and so
\[
|F(r)|\leq C\frac{[\log(e+r)]^{2\chi(\delta)+1-H(\delta)}}{(1+r)^{1-2\delta^+}}X_\delta^2,\quad r\geq 1.
\]
On the other hand, for \(0<r\leq 1\),
\[
\aligned
|F(r)|\leq & \frac{3}{r^2}\int_0^rs^2(f(u,r)-f(u,s))ds\\
\leq & \frac{1}{r^2}\int_0^rs^2(e^{CX_\delta^2}-1) ds
\lesssim (e^{CX_\delta^2}-1)r
\lesssim \frac{[\log(e+r)]^{2\chi(\delta)+1-H(\delta)}}{(1+r)^{1-2\delta^+}}X_\delta^2,
\endaligned
\] 
where here the implicit constant depends on \(\gamma\) and \(\delta\).
Therefore,
\[
|F(r)|\leq C\frac{[\log(e+r)]^{2\chi(\delta)+1-H(\delta)}}{(1+r)^{1-2\delta^+}}X_\delta^2,\ \forall r>0.
\]
Now, using \eqref{eq12}, we see that 
\[
\|\partial_rf\|_{L_U^\infty L_r^{\infty,2-\delta}}\leq CX_\delta^2,
\] 
hence
\[
\frac{1}{r}|f-\bar{f}|\leq C\frac{\log^{\chi(\delta)}(e+r)}{(1+r)^{2-\delta^+}}\|\partial_rf\|_{L_U^\infty L_r^{\infty,2-\delta}}
\leq C\frac{[\log(e+r)]^{2\chi(\delta)+1-H(\delta)}}{(1+r)^{1-2\delta^+}}X_\delta^2
\]
as well as
\[
r^2\partial_rf\leq C\frac{r^3\log^{2\chi(\delta)}(e+r)}{(1+r)^{4-2\delta^+}}X_\delta^2\leq C\frac{[\log(e+r)]^{2\chi(\delta)+1-H(\delta)}}{(1+r)^{1-2\delta^+}}X_\delta^2.
\]
Combining these estimates with \eqref{eq8} and  \eqref{diff-J} we get
\[
|J|\leq \frac{3}{2r}|f-\bar{f}|+\frac{3}{2}|F(r)|+\frac{|1-3r^2|}{2}\partial_rf\leq C\frac{[\log(e+r)]^{2\chi(\delta)+1-H(\delta)}}{(1+r)^{1-2\delta^+}}X_\delta^2.
\]
Finally we turn our attention to \(\partial_rJ\) which, using \eqref{diff-J}, can be written as 
\begin{equation}
\aligned
\partial_rJ =& \frac{3}{2r}\left( \partial_r f-\partial_r\bar f- \frac{f-\bar f}{r} -2 F(r)-r^2\partial_rf \right) +\frac{3r^2+1}{4r^2}f(h-\bar{h})^2\\
&+\frac{3r^2-1}{4r}\big( \partial_rf(h-\bar{h})^2+2f(h-\bar{h})(\partial_rh-\partial_r\bar{h})\big).
\endaligned
\end{equation}
Then using the estimates we've seen above we get estimate \eqref{eq24}, for \(r\geq 1\) and \(\gamma\) small enough.
\end{proof}
\subsection{Estimates along the characteristics}
\label{sectionChar}
We now consider the estimates along the characteristics of \eqref{eq3}, which we denote by \(\chi(u)=\chi(u;u_1,r_1)=(u,r(u;u_1,r_1))\), where \(r\) is the unique solution to
\begin{equation}\label{eq19}
\frac{dr}{du}=-\frac{1}{2}\tilde{f}(u,r),
\end{equation}
such that \(r(u_1)=r_1\). To obtain estimates for these solutions we can estimate \(\tilde{f}\). Using \eqref{eq7} and \eqref{eq16} we have
\begin{equation}\label{eq18}
1-(1+\varepsilon_{X_\delta})r^2\leq \tilde{f}(u,r)\leq 1+\varepsilon_{X_\delta}-r^2.
\end{equation}
From this observation we can easily show the next useful result:
\begin{lemma}
\label{lemma2.2}
Consider a characteristic \(\chi(u)=(u,r(u;u_1,r_1))\) with \(r_1\geq R\). For \(X_\delta\) small enough and \(R\) large enough we have that 
\begin{equation}\label{eq17}
\frac{2(1-\varepsilon_{X_{\delta,R}})}{r(u)^2}\frac{dr}{du}\leq 1=\frac{2}{-\tilde{f}(u,r(u))}\frac{dr}{du}\leq \frac{2(1+\varepsilon_{X_{\delta,R}})}{r(u)^2}\frac{dr}{du},\text{ for }u\in[u_R,u_1],
\end{equation}
where \(u_R=\max(\{u\in [0,u_1]: r(u)=R\}\cup \{0\})\) and where~\footnote{The use of the notation \(\varepsilon_{X_{\delta,R}}\) mirrors that of \(\varepsilon_{X_\delta}\).} \(\varepsilon_{X_{\delta,R}}>0\) represents a quantity which goes to zero as \(X_\delta\rightarrow0\) and \(R\rightarrow \infty\). Moreover, the expression \(\varepsilon_{X_{\delta,R}}\) on the left hand side of \eqref{eq17} can be chosen so that 
$r_1\geq (1-\varepsilon_{X_{\delta,R}})(1+r_1)$.
\end{lemma}
Now we recall that in  \cite{CostaProblem} it was established:
\begin{lemma}
\label{lemma2.3}
Consider a characteristic \(\chi(u)=(u,r(u; u_1,r_1))\), with \(r_1>R\). For \(R>0\) large enough we have
\[
r(u)\geq (1-\varepsilon_{X_\delta})\coth\left(\frac{1+\varepsilon_{X_\delta}}{2}(c^--u)\right), \forall u\leq u_1,
\]
where \(c^-\) is chosen such that 
\[
r_1=(1-\varepsilon_{X_\delta})\coth\left(\frac{1+\varepsilon_{X_\delta}}{2}(c^--u_1)\right).
\]
\end{lemma}
Using the previous lemmas we will now show a generalisation of Lemma 2  of \cite{Costa-Mena} by including an extra \(\log\) factor. This result will be crucial in what follows as it represents a gain of a unit power of polynomial decay for our estimates. 
\begin{prop}\label{prop2.3}
Let $m>0$, $k\ge 0$ and \(0< p<2m-1\). Then, there exist \(\gamma, C>0\) such that, if \(X_\delta<\gamma\), then
\begin{equation}\label{eq23}
\int_0^{u_1}\frac{\log^k(e+r(u))}{(1+r(u))^p}e^{m\int_u^{u_1}G(s,r(s;u_1,r_1))ds}du\leq C \frac{\log^k(e+r_1)(1+u_1)}{(1+r_1)^{p+1}},\text{ for }u_1\leq U.
\end{equation}
\end{prop}
\begin{proof}
The proof is analog to that of Lemma 2 of \cite{Costa-Mena} so we just summarize some differences. Since the function
\[
\frac{\log^k(e+r)}{(1+r)^p}
\]
is continuous on \([0,\infty[\) with a finite limit as \(r\rightarrow \infty\), it must be bounded over \([0,\infty[\). So we have, for \(\gamma\) small enough,
\[
\int_0^{u_1}\frac{\log^k(e+r(u))}{(1+r(u))^p}e^{m\int_u^{u_1}Gds}du\leq \int_0^{u_1}\frac{\log^k(e+r(u))}{(1+r(u))^p}du\lesssim u_1.
\]
Then, the estimate \eqref{eq23} follows for \(r_1\leq R\), for a sufficiently large $R$. 

In the case \(r_1\geq R\) we use similar estimates as in Lemma 2 of \cite{Costa-Mena} together with  Lemma \ref{lemma2.2} and Lemma \ref{lemma2.3} to get
\[
\begin{split}
\int_0^{u_R}\frac{\log^k(e+r(u))}{(1+r(u))^p}e^{m\int_u^{u_1}Gds}du&\leq \frac{1+\varepsilon_{X_{\delta,R}}}{(1+r_1)^{2m(1-\varepsilon_{X_\delta})}}\int_0^{u_R}\log^k(e+r(u))(1+r(u))^{2m(1-\varepsilon_{X_{\delta,R}})-p}du\\
&\lesssim \log^k(e+r)(1+R)^{2m(1-\varepsilon_{X_{\delta,R}})-p}\frac{(1+r_1)^{p+1}}{(1+r_1)^{2m(1-\varepsilon_{X_\delta})}}\frac{u_R}{(1+r_1)^{p+1}}\\
&\lesssim \frac{1+u_1}{(1+r_1)^{p+1}},
\end{split}
\]
with $u_R$ as defined in Lemma \ref{lemma2.2}. Finally,
\[
\begin{split}
\int_{u_R}^{u_1}\frac{\log^k(e+r)}{(1+r)^p}e^{m\int_u^{u_1}Gds}du&\leq \frac{2(1+\varepsilon_{X_{\delta,R}})}{(1-\varepsilon_{X_{\delta,R}})(1+r_1)^{2m(1-\varepsilon_{X_{\delta,R}})}}\int_{r(u_R)}^{r_1}\log^k(e+r)(1+r)^{2m(1-\varepsilon_{X_{\delta,R}})-p-2}dr\\
&\lesssim \frac{\log^k(e+r_1)}{(1+r_1)^{2m(1-\varepsilon_{X_{\delta,R}})}}(1+r_1)^{2m(1-\varepsilon_{X_{\delta,R}})-p-1}\\
&\lesssim \frac{\log^k(e+r_1)(1+u_1)}{(1+r_1)^{p+1}},
\end{split}
\]
which concludes the proof.
\end{proof}
\section{Global existence of classical solutions}
\label{global-sec}
In this section we generalise the global existence results of  \cite{Costa-Mena} which were valid only for $0<\delta<1/2$. The strategy of the proofs are similar but harder since new $\log$ terms appear in some crucial expressions and sharper estimates are necessary. Furthermore, new estimates for $\partial_r J$ are necessary in order to close the iteration method.
\subsection{Local existence results}
We begin by stating a local existence theorem whose proof can be found in appendix A of \cite{Costa-Mena}.
\begin{thm}\label{thm3.1}
Given \(\tau_0>0\) and \(h_0\in C^k([0,\tau_0]), k\in \N\), there exists a positive
\[
\tau=\tau\left(\sup_{0\leq r\leq \tau_0}|h_0(r)|,\ \sup_{0\leq r\leq \tau_0}|h_0'(r)|\right)\leq \tau_0
\]
and a unique solution \(h\in C^k([0,\tau]^2)\) to
\[
\begin{cases}
Dh=G(h-\bar{h})\\
h(0,r)=h_0(r).
\end{cases}
\]
\end{thm}
We will now use this theorem together with the estimates we saw in the previous section to prove the existence of a solution to the Einstein-massless scalar field system that is local in Bondi time but global in radius. The main result of this subsection is:
\begin{thm}
\label{thm3.2}
Let \(-1\leq\delta<1/2\) and \(k\in \N\). Suppose that \(h_0\in C^{k+1}([0,\infty[)\cap L^\infty([0,\infty[)\) satisfies 
\[
\|h_0'\|_{L_r^{\infty,2-\delta}}<\infty\text{ and }\|h_0''\|_{L_r^{\infty,3-\delta}}<\infty.
\]
Then, there exists \(\gamma>0\) independent of \(h_0\) such that, when
\[
\|h_0'\|_{L_r^{\infty,2-\delta}}\leq \gamma,
\]
there exists \(U=U(\gamma, \delta)>0\) and a unique solution \(h\in C^{k+1}([0,U]\times [0,\infty[)\) to
\begin{equation}
\aligned
\label{xcvbn}
&Dh=G(h-\bar{h})\\
&h(0,r)=h_0(r).
\endaligned
\end{equation}
Moreover, \(\|\partial_rh\|_{L_U^\infty L_r^{\infty,2-\delta}}\) can be made as small as we want by decreasing \(\gamma\).
\end{thm}
The proof of this theorem is summarised at the end of this section and follows from the coming Proposition \ref{prop3.1}, Proposition \ref{prop3.2} and from recycling some parts of the proof of Theorem 2 in \cite{Costa-Mena}. 
The proof is based on showing that a certain sequence \((h_n)\) defined below contracts in \(L_U^\infty L_r^{\infty}\). 
We now explain how we construct this sequence. Following \cite{Costa-Mena} we define the functions:
\begin{align}
&f_n=\exp\left(\frac{1}{r}\int_0^r\frac{(h_n-\bar{h}_n)^2}{s}ds\right)\\
&\tilde{f}_n=\frac{1}{r}\int_0^r(1-3s^2)f_n(u,s)ds\\
&G_n=\frac{1}{2}\partial_r\tilde{f}_n\\
&J_n=G_n-r\partial_rG_n.
\end{align}
We consider also the operator
\[
D_n=\frac{\partial}{\partial u}-\frac{\tilde{f}_n}{2}\frac{\partial}{\partial r}
\]
and its corresponding characteristics \(\chi_n(u)=(u,r_n(u;u_1,r_1))\), where \(r_n\) is the unique solution to
\[
\frac{dr_n}{du}=-\frac{1}{2}\tilde{f}_n(u,r_n(u))
\]
with \(r_n(u_1)=r_1\). Now we define \(h_n\) as
\begin{equation}\label{eq44}
h_n(u,r)=h_U(u,0)+\int_0^rw_n(u,s)ds,
\end{equation}
where \(h_U\) is the unique solution to \eqref{xcvbn} which, according to Theorem \ref{thm3.1}, exists for small enough \(U\), and  where \(w_n\) is the sequence defined recursively by setting \(w_1(u,r)=h_0'(r)\) and defining \(w_{n+1}\) to be the solution to
\begin{equation}\label{eq29}
\begin{cases}
D_nw_{n+1}=2G_nw_{n+1}-J_n\frac{h_n-\bar{h}_n}{r}\\
w_{n+1}(0,r)=h_0'(r).
\end{cases}
\end{equation}
So, we will start by showing that for small enough initial data, the sequences \((h_n), (w_n)\) and \((\partial_r w_n)\) are bounded uniformly in \(n\), in appropriate spaces. First we observe that by integrating \eqref{eq29} the following relations hold:
\begin{equation}
\label{lemma3.1}
\aligned
w_{n+1}(u_1,r_1)&=h_0'(r_n(0))e^{\int_0^{u_1}2G_nds}-\int_0^{u_1}\frac{J_n(u,r_n(u))(h_n-\bar{h}_n)(u,r_n(u))}{r_n(u)}e^{\int_u^{u_1}2G_nds}du\\
\partial_rw_{n+1}(u_1,r_1)&=h_0''(r_n(0))e^{\int_0^{u_1}3G_nds}+2\int_0^{u_1}\partial_rG_nw_{n+1}e^{\int_u^{u_1}3G_nds}du\\
&\quad-\int_0^{u_1}\partial_r J_n\frac{h_n-\bar{h}_n}{r_n}e^{\int_u^{u_1}3G_nds}du+\int_0^{u_1}\frac{J_n}{r_n}\left[2\frac{h_n-\bar{h}_n}{r_n}-w_n\right]e^{\int_u^{u_1}3G_nds}du.
\endaligned
\end{equation}
The next proposition generalises Lemma 3 of \cite{Costa-Mena}, nevertheless its proof is more delicate as now the integrals involving $J$ have to be estimated in different regions of the characteristics and, furthermore, shaper estimates for $\partial_r J$ are needed due to the more general decay considered for the initial data. 
\begin{prop}
\label{prop3.1}
Under the same assumptions as in Theorem \ref{thm3.2}, there exists some \(x_0'>0\) such that, if
\[
\|h_0'\|_{L_r^{\infty,2-\delta}}\leq x_0',
\]
then there are constants \(C, x',x''>0\) for which
\begin{equation}\label{eq30}
\aligned
 \|h_n \|_{L_U^\infty L_r^\infty}&\leq \sup_{0\leq u\leq U}|h_U(u,0)|+Cx' \\
 \|w_n  \|_{L_U^\infty L_r^{\infty,2-\delta}}&\leq x'\\
 \|\partial_r  w_n\|_{L_U^\infty L_r^{\infty,3-\delta}}&\leq x'',
\endaligned
\end{equation}
uniformly on \(n\in \N\). Furthermore, \(x'\) can be made as small as we want by decreasing \(x_0'\) and similarly \(x''\) can be made as small as needed by decreasing both \(x_0'\) and \(\|h_0''\|_{L_r^{\infty,3-\delta}}\).
\end{prop}
\begin{proof}
The proof is done by induction on \(n\). For \(n=1\) we have \(w_1(u,r)=h_0'(r)\), so \(\|w_1\|_{L_U^\infty L_r^{\infty,2-\delta}}=\|h_0'\|_{L_r^{\infty,2-\delta}}=:x_0'\). Also, \(\partial_rw_1(u,r)=h_0''(r)\) and thus \(\|\partial_rw_1\|_{L_U^\infty L_r^{\infty,3-\delta}}=\|h_0''\|_{L_r^{\infty,3-\delta}}=:x_0''\). If we set
\[
b_0=\sup_{0\leq u\leq U}|h_U(u,0)|,
\]
we also have
\[
|h_1(u,r)|\leq b_0+\int_0^r(1+s)^{2-\delta}|w_1(u,s)|\frac{1}{(1+s)^{2-\delta}}ds\leq b_0+Cx_0'\;.
\]
Therefore the result holds for \(n=1\). Now we suppose that the result holds for some \(n\in \N\), i.e. we have the estimates 
\[
\begin{array}{l}
\|h_n\|_{L_U^\infty L_r^\infty}\leq b_0+Cx';\vspace{5pt}\\
\|w_n\|_{L_U^\infty L_r^{\infty,2-\delta}}\leq x';\vspace{5pt}\\
\|\partial_r w_n\|_{L_U^\infty L_r^{\infty,3-\delta}}\leq x'',
\end{array}
\]
for some constants \(x', x''\) to be fixed below. First we note that
\begin{align}
|h_{n+1}(u,r)|&\leq b_0+\int_0^r|w_{n+1}(u,s)|ds
\\
&\leq b_0+\|w_{n+1}\|_{L_U^\infty L_r^{\infty,2-\delta}}\int_0^r\frac{1}{(1+s)^{2-\delta}}ds
\\
&\leq b_0+C\|w_{n+1}\|_{L_U^\infty L_r^{\infty,2-\delta}},
\end{align}
so the estimate for \(h_{n+1}\) follows from the estimate for \(w_{n+1}\). Now we turn to the estimation of \(\|w_{n+1}\|_{L_U^\infty L_r^{\infty,2-\delta}}\). Using (\ref{lemma3.1}) we see that 
\[
|w_{n+1}(u_1,r_1)|\leq A_1+A_2,
\]
where
\begin{align}
&A_1=|h_0'(r_n(0))|e^{\int_0^{u_1}2G_nds};\vspace{10pt}\\
&A_2=\int_0^{u_1}\frac{|J_n(h_n-\bar{h}_n)|}{r_n}e^{\int_u^{u_1}2G_nds}du.
\end{align}
We now estimate these individually. For \(A_1\) we can apply Proposition \ref{prop2.3}, in view of the induction hypothesis, possibly by decreasing the value of \(x'\), and then  use \eqref{eq23} with \(\eta\) small enough to obtain
\[
\begin{split}
(1+r_1)^{2-\delta}A_1&\lesssim (1+r_1)^{2-\delta}|h_0'(r_n(0))|\left(\frac{1+r_n(0)}{1+r_1}\right)^{4-\eta}
\\
&\lesssim (1+r_n(0))^{2-\delta}|h_0'(r_n(0))|\left(\frac{1+r_n(0)}{1+r_1}\right)^{2-\eta+\delta}\\
&\lesssim \|h_0'\|_{L_r^{\infty,2-\delta}}\left(\frac{1+r_n(0)}{1+r_1}\right)^{2-\eta+\delta}\lesssim x_0'\left(\frac{1+r_n(0)}{1+r_1}\right)^{2-\eta+\delta}.
\end{split}
\]
Noting that for \(r_n(0)\) large, \(r(u)\) is increasing in \(u\) we see that the sequence \((1+r_n(0))/(1+r_1)\) is bounded uniformly on \(n\) and \(r_1\), so
\[
(1+r_1)^{2-\delta}A_1\lesssim x_0'\;.
\]
To estimate \(A_2\) we use Proposition \ref{prop1} and Proposition \ref{prop2.3} to get
\[
\begin{split}
(1+r_1)^{2-\delta}A_2&\lesssim(1+r_1)^{2-\delta} (x')^3\int_0^{u_1}\frac{[\log(e+r_n)]^{3\chi(\delta)+1-H(\delta)}}{(1+r_n)^{3-3\delta^+}}e^{\int_u^{u_1}2G_nds}du\\
&\lesssim (1+r_1)^2(x')^3\frac{[\log(e+r_1)]^{3\chi(\delta)+1-H(\delta)}}{(1+r_1)^{4-3\delta^+-\eta}}(1+u_1)\lesssim (x')^3.
\end{split}
\]
In this way we have
\begin{equation}\label{eq33}
\|w_{n+1}\|_{L_U^\infty L_r^{\infty,2-\delta}}\lesssim x_0'+(x')^3.
\end{equation}
We now turn our attention to estimating the norm of \(\partial_rw_{n+1}\). To this end we use \eqref{lemma3.1} to obtain
\[
|\partial_rw_{n+1}(u_1,r_1)|\leq B_1+B_2+B_3+B_4,
\]
where
\begin{equation}
\aligned
B_1&=|h_0''(r_n(0))|e^{\int_0^{u_1}3G_nds}\\
B_2&=2\int_0^{u_1}|\partial_rG_n||w_{n+1}|e^{\int_u^{u_1}3G_nds}du\\
B_3&=\int_0^{u_1}|\partial_rJ_n|\frac{|h_n-\bar{h}_n|}{r_n}e^{\int_u^{u_1}3G_nds}du\\
B_4&=\int_0^{u_1}\frac{|J_n|}{r_n}\left[2\frac{|h_n-\bar{h}_n|}{r_n}+|w_n|\right]e^{\int_u^{u_1}3G_nds}du.
\endaligned
\end{equation}
At this point we estimate these individually. Note again that, using the induction hypothesis, we may apply Proposition \ref{prop1} and Proposition \ref{prop2.3}, reducing the value of \(x'\) if needed. First we estimate \(B_1\) as
\[
\begin{split}
(1+r_1)^{3-\delta}B_1&=\frac{(1+r_1)^{3-\delta}}{(1+r_n(0))^{3-\delta}}(1+r_n(0))^{3-\delta}|h_0''(r_n(0))|e^{\int_0^{u_1}3G_nds}\\
&\lesssim x_0''\frac{(1+r_1)^{3-\delta}}{(1+r_n(0))^{3-\delta}}\left(\frac{1+r_n(0)}{1+r_1}\right)^{6-\eta}\\
&\lesssim x_0''\left(\frac{1+r_n(0)}{1+r_1}\right)^{3-\eta+\delta}\lesssim x_0''.
\end{split}
\]
Now before estimating \(B_2\) we first observe some facts: 
\begin{enumerate}
\item Since \(|f_n|\lesssim 1\), we have that
\[
|\partial_rf_n|=\left|\frac{f_n(h_n-\bar{h}_n)^2}{2r_n}\right|\lesssim \frac{r_n\log^{2\chi(\delta)}(e+r_n)}{(1+r_n)^{4-2\delta^+}}(x')^2.
\]
\item From Proposition \ref{prop1} we know that \(|G_n|\lesssim r_n\), so
\[
\frac{|G_n|}{r_n}\lesssim 1.
\]
\item Using \eqref{diff-J},
\[
\frac{|J_n|}{r_n}\leq 3\frac{|G_n|}{r_n}+3|f_n|+\frac{1+3r_n^2}{2r_n}|\partial_rf_n|\lesssim 1+1+\frac{\log^{2\chi(\delta)}(e+r_n)}{(1+r_n)^{2-2\delta^+}}(x')^2\lesssim 1.
\]
\item Putting these estimates together and noting that \(\partial_rG_n=(G_n-J_n)/r_n\) we obtain
\[
|\partial_rG_n||w_{n+1}|=(1+r_n)^{2-\delta}|w_{n+1}|\frac{|\partial_rG_n|}{(1+r_n)^{2-\delta}}\lesssim \frac{x_0'+(x')^3}{(1+r_n)^{2-\delta}}.
\]
\end{enumerate}
Now we can use Proposition \ref{prop2.3} to estimate \(B_2\) as
\[
\begin{split}
(1+r_1)^{3-\delta}B_2&\lesssim (1+r_1)^{3-\delta}(x_0'+(x')^3)\int_0^{u_1}\frac{1}{(1+r_n)^{2-\delta}}e^{\int_u^{u_1}3G_nds}du\\
&\lesssim (1+r_1)^{3-\delta}(x_0'+(x')^3)\frac{1+u_1}{(1+r_1)^{3-\delta}}\lesssim x_0'+(x')^3.
\end{split}
\]
Next we estimate \(B_4\). To do this we split the integral in two parts. 

Let \(\Omega_{\leq 1}=\{u\in [0,u_1]: r_n(u)\leq 1\}\) and \(\Omega_{>1}=\{u\in [0,u_1]: r_n(u)>1\}\). If \(u\in \Omega_{\leq 1}\) we have
\[
\begin{split}
\frac{|J_n|}{r_n}\frac{|h_n-\bar{h}_n|}{r_n}+\frac{|J_n|}{r_n}|w_n|&\lesssim \frac{\log^{\chi(\delta)}(e+r_n)}{(1+r_n)^{2-\delta^+}}x'+\frac{x'}{(1+r_n)^{2-\delta}}\\
&\lesssim \frac{\log^{\chi(\delta)}(e+r_n)(1+r_n)^{\delta^+-\delta}}{(1+r_n)^{2-\delta}}x'+\frac{x'}{(1+r_n)^{2-\delta}}\lesssim \frac{x'}{(1+r_n)^{2-\delta}}.
\end{split}
\]
On the other hand, when \(u\in \Omega_{>1}\) we will use estimate \eqref{eq15} for \(J_n\). This is in contrast to what is done in \cite{Costa-Mena}, where such care was unnecessary. In this case we obtain,
\[
\frac{|J_n|}{r_n}\frac{|h_n-\bar{h}_n|}{r_n}+\frac{|J_n|}{r_n}|w_n|\lesssim \frac{[\log(e+r_n)]^{3\chi(\delta)+1-H(\delta)}}{(1+r_n)^{4-3\delta^+}}(x')^3.
\]
Equipped with these estimates and using Proposition \ref{prop2.3}, we see that
\[
\begin{split}
(1+r_1&)^{3-\delta}B_4=(1+r_1)^{3-\delta}\left(\int_{\Omega_{\leq 1}}\frac{|J_n|}{r_n}\left[2\frac{|h_n-\bar{h}_n|}{r_n}+|w_n|\right]e^{\int_u^{u_1}3G_nds}du\right.\\
&\left.+\int_{\Omega_{> 1}}\frac{|J_n|}{r_n}\left[2\frac{|h_n-\bar{h}_n|}{r_n}+|w_n|\right]e^{\int_u^{u_1}3G_nds}du\right)\\
&\lesssim (1+r_1)^{3-\delta}\left(x'\int_0^{u_1}\frac{1}{(1+r_n)^{2-\delta}}e^{\int_u^{u_1}3G_nds}du+(x')^3\int_0^{u_1}\frac{[\log(e+r_n)]^{3\chi(\delta)+1-H(\delta)}}{(1+r_n)^{4-3\delta^+}}e^{\int_u^{u_1}3G_nds}du\right)\\
&\lesssim (1+r_1)^{3-\delta}\left(x'\frac{1+u_1}{(1+r_1)^{3-\delta}}+(x')^3\frac{[\log(e+r_1)]^{3\chi(\delta)+1-H(\delta)}}{(1+r_1)^{5-3\delta^+}}(1+u_1)\right)\lesssim x'+(x')^3. 
\end{split}
\]
Finally, we estimate \(B_3\) using an analogous splitting as above. Note that we have
\[
\begin{split}
|\partial_rJ_n|&=\left|3\partial_rG_n+3f_n+3r\partial_rf_n-\partial_rf_n\frac{1-3r^2}{4r}(h_n-\bar{h}_n)^2+f_n\frac{1+3r^2}{4r^2}(h_n-\bar{h}_n)^2\right.\\
&\left.-f_n\frac{1-3r^2}{2r}(h_n-\bar{h}_n)\left(w_n-\frac{h_n-\bar{h}_n}{r}\right)\right|\lesssim 1.
\end{split}
\]
So, for \(u\in \Omega_{\leq 1}\) we have the estimate
\[
|\partial_rJ_n|\frac{|h_n-\bar{h}_n|}{r_n}\lesssim \frac{\log^{\chi(\delta)}(e+r_n)}{(1+r_n)^{2-\delta}}x'\lesssim \frac{\log(e+1)}{(1+r_n)^{2-\delta}}x'\lesssim \frac{x'}{(1+r_n)^{2-\delta}}\;.
\]
On the other hand, if \(u\in \Omega_{>1}\) we can use \eqref{eq24}. We remark that the estimate \eqref{eq24} is sharper than the estimate used in \cite{Costa-Mena}, when \(r\geq 1\). While in \cite{Costa-Mena} this wasn't necessary, here it is crucial. We then have
\[
|\partial_rJ_n|\frac{|h_n-\bar{h}_n|}{r_n}\lesssim \frac{[\log(e+r_n)]^{3\chi(\delta)+1-H(\delta)}}{(1+r_n)^{4-3\delta^+}}(x')^3.
\]
With these estimates we get
\[
\begin{split}
(1+r_1&)^{3-\delta}B_3=(1+r_1)^3\left(\int_{\Omega_{\leq 1}}|\partial_rJ_n|\frac{|h_n-\bar{h}_n|}{r_n}e^{\int_u^{u_1}3G_nds}du+\int_{\Omega_{>1}}|\partial_rJ_n|\frac{|h_n-\bar{h}_n|}{r_n}e^{\int_u^{u_1}3G_nds}du\right)\\
&\lesssim (1+r_1)^{3-\delta}\left(x'\int_0^{u_1}\frac{1}{(1+r_n)^{2-\delta}}e^{\int_{u}^{u_1}3G_nds}du+(x')^3\int_0^{u_1}\frac{[\log(e+r_n)]^{3\chi(\delta)+1-H(\delta)}}{(1+r_n)^{4-3\delta^+}}e^{\int_u^{u_1}3G_nds}du\right)\\
&\lesssim (1+r_1)^{3-\delta}\left(x'\frac{1+u_1}{(1+r_1)^{3-\delta}}+(x')^3\frac{[\log(e+r_1)]^{3\chi(\delta)+1-H(\delta)}(1+u_1)}{(1+r_1)^{5-3\delta^+}}\right)\lesssim x'+(x')^3.
\end{split}
\]
Putting these estimates together we see that
\begin{equation}\label{eq38}
\|\partial_rw_{n+1}\|_{L_U^\infty L_r^{\infty,3-\delta}}\lesssim x_0''+x_0'+x'+(x')^3.
\end{equation}
Denote by \(C_1\) the constant implicit in \eqref{eq33} and by \(C_2\) the one implicit in \eqref{eq38}. Since these constants do not depend on \(n\) we can choose \(x_0'\) and \(x'\) to be small enough such that \(C_1(x_0'+(x')^3)\leq x'\). Moreover, these can be chosen so that \(x_0'\leq x'\) and \(x'\) can be made smaller by decreasing \(x_0'\). With these choices we now put \(x'':=\max\{C_2(x_0''+x_0'+x'+(x')^3), x_0''\}\). So we get
\[
\begin{array}{l}
\|h_{n+1}\|_{L_U^\infty L_r^\infty}\leq \sup_{0\leq u\leq U}|h_U(u,0)|+Cx';\vspace{5pt}\\
\|w_{n+1}\|_{L_U^\infty L_r^{\infty,2-\delta}}\leq x';\vspace{5pt}\\
\|\partial_r w_{n+1}\|_{L_U^\infty L_r^{\infty,3-\delta}}\leq x'',
\end{array}
\]
thus concluding the induction step.
\end{proof}
\begin{proof}[Proof of Theorem \ref{thm3.2}]
From Proposition \ref{prop3.2} in the appendix we see that \((h_n)_n\) defined above converges uniformly to some continuous function \(h:[0,U]\times [0,\infty[\rightarrow \R\). Using estimates similar to those we've already deduced in Proposition \ref{prop3.1} one can show that the sequences \((f_n), (\tilde{f}_n), (G_n)\) and \((J_n)\) converge uniformly to functions \(f, \tilde{f}, G,J\), where the convergence \(f_n\rightarrow f\) is over \([0,U]\times [0,\infty[\), whereas the others converge uniformly over all intervals of the form \([0,U]\times [0,R]\). Moreover, the characteristics \(r_n(\ \cdot\ ; u_1,r_1)\) also converge uniformly over \([0,U]\times [0,R]\) to characteristics \(r(\ \cdot\ ;u_1,r_1)\) of \(D=\partial_u-\tilde{f}\partial_r/2\). We can now use a uniqueness result and methods analogous to those of Proposition 2 of \cite{Costa-Mena} to finish the proof of Theorem \ref{thm3.2}.
\end{proof}
%
\subsection{Global existence of the solutions in time and radius}
%
\begin{thm}
\label{thm3.3}
Let \(-1<\delta<1/2\) and suppose that \(h_0\in C^{k+1}([0,\infty[)\cap L^\infty([0,\infty[), k\in \N,\) is such that \(h_0'\in L^{\infty,2-\delta}([0,\infty[)\) and \(h_0''\in L^{\infty,3-\delta}([0,\infty[)\). In this case, there exists some \(\tilde{x}_0>0\) such that, if 
\[
\|h_0'\|_{L_r^{\infty,2-\delta}}\leq \tilde{x}_0
\]
then the problem
\begin{equation}
\label{asdfg}
\begin{dcases}
Dh=G(h-\bar{h});\\
h(0,r)=h_0(r)
\end{dcases}
\end{equation}
has a unique solution \(h\in C^{k+1}([0,\infty[\times [0, \infty[)\). 
\end{thm}
\begin{proof}
Taking into account the results of the previous section, the proof of this theorem now follows from using a strategy similar to the proof of Theorem 3 of \cite{Costa-Mena} with small modifications. So we just give a brief sketch of the proof while providing results that will be used in the sequel.

Let \(-1<\delta<1/2\) and consider \(h_0\) in the same conditions as in Theorem \ref{thm3.2}. 

According to Theorem \ref{thm3.2} there is some \(\gamma>0\) such that, if \(\|h_0'\|_{L_r^{\infty,2}}\leq \gamma\), then the system \eqref{asdfg} 
has a unique solution \(h\in C^{k+1}([0,U]\times [0,\infty[)\), for some \(U>0\). Now consider some \(\tilde{x}_0<\gamma\) and suppose that \(\|h_0'\|_{L_r^{\infty,2-\delta}}\leq \tilde{x}_0\). Then we obtain some \(\tilde{U}=\tilde{U}(\tilde{x}_0)\) coming from Theorem \ref{thm3.2}. Here we define \(U^*\in [\tilde{U}, \infty]\) to be the maximal time of existence of solutions. So there is some \(h\in C^{k+1}([0,U^*[\times [0,\infty[)\) satisfying \eqref{asdfg}.

Note that, if we take the derivative with respect to \(r\) of the differential equation in \eqref{asdfg}, then consider it along the characteristics \((u,r(u;u_1,r_1))\), multiply it by an appropriate integrating factor and then integrate in \(u\) we obtain
\[
\partial_rh(u_1,r_1)=h_0'(r(0))e^{\int_0^{u_1}2G(s,r(s))ds}-\int_0^{u_1}\frac{J}{r}(h-\bar{h})e^{\int_u^{u_1}2Gds}du,
\]
where $(u_1,r_1)\in [0,U^*[\times [0,\infty[.$
The proof of global existence of solutions is now based on providing estimates on the energy function
\[
\mcal{E}_R(u)=\sup_{r\in[0,R]}|(1+r)^{2-\delta}\partial_rh(u,r)|,
\]
for a fixed \(R\gg 1\). Since the supremum is being taken over a compact set, \(\mcal{E}_R\) is continuous. In particular, the set 
\[
\mcal{U}_R=\{u_1\in[0,U^*[: \sup_{u\in [0,u_1]}\mcal{E}_R(u)\leq x'\}
\]
is closed, where \(x'\in ]\tilde{x}_0,\gamma[\) is  specified during the proof (see \cite{Costa-Mena}). It is also clearly non-empty (\(0\in \mcal{U}_R\)) by our assumption on \(\|h_0'\|_{L_r^{\infty,2-\delta}}\). The goal of the proof is to show that this set is also open (in the relative topology) and therefore equal to $[0,+\infty[$. To do that we show that for \(u_1\in \mcal{U}_R\) we can improve the estimate \(\mcal{E}_R(u_1)\leq x'\). An important point is that for \(u_1\in \mcal{U}_R\), using an argument analogous to that of Lemma \ref{lemma2.3}, one can show that there is some 
\begin{equation}
\label{rc}
r_c^-=1-\varepsilon_{x'}
\end{equation} 
such that if \(r_1>r_c^-\) then \(r(u;u_1,r_1)>r_c^-\) for \(u\leq u_1\), and if \(r_1\leq r_c^-\)
\[
r(u)\geq (1-\varepsilon_{x'})\tanh\left(\frac{1+\varepsilon_{x'}}{2}(c^--u)\right).
\]
which allows us to provide estimates for small values of $r$. 
From this we obtain
\[
\begin{split}
-\int_u^{u_1}r(v)dv&\leq -(1-\varepsilon_{x'})\int_u^{u_1}\tanh\left(\frac{1+\varepsilon_{x'}}{2}(c^--v)\right)dv \leq 2(1-\varepsilon_{x'})\log\left(2e^{\frac{1+\varepsilon_{x'}}{2}(u-u_1)}\right).
\end{split}
\]
So, for \(r_1\leq r_c^-\) we have that
\begin{equation}
\label{exp-estimate}
e^{\int_u^{u_1}2G(s,r(s))ds}\leq e^{-2\int_u^{u_1}(1-\varepsilon_{x'})rds}\lesssim e^{2(1-\varepsilon_{x'})(u-u_1)}.
\end{equation}
In this way we see that, for \(r_1\leq r_c^-\),
\[
\begin{split}
|(1+r_1)^{2-\delta}\partial_rh(u_1,r_1)|&\leq (1+r_1)^{2-\delta}|h_0'(r(0))|e^{\int_0^{u_1}2Gds}+(1+r_1)^{2-\delta}\int_0^{u_1}\frac{|J|}{r}|h-\bar{h}|e^{\int_u^{u_1}2Gds}du\\
&\lesssim \mcal{E}_R(0)e^{-2(1-\varepsilon_{x'})u_1}+x'\int_0^{u_1}\frac{[\log(e+r)]^{3\chi(\delta)+1-H(\delta)}}{(1+r)^{3-3\delta^+}}\mcal{E}_R(u)e^{2(1-\varepsilon_{x'})(u-u_1)}du\\
&\lesssim \mcal{E}_R(0)e^{-2(1-\varepsilon_{x'})u_1}+x'\int_0^{u_1}\mcal{E}_R(u)e^{2(1-\varepsilon_{x'})(u-u_1)}du.
\end{split}
\]
where we used the fact that \(r(u)\lesssim 1\) for \(u\leq u_1<U^*\) in the case when \(r_1<r_c^-\), which follows from estimate (30) of \cite{CostaProblem}. In this way we can take \(R\) large enough so that \(r(u)\leq R,\ 0\leq u\leq u_1\). Also, in the second step we used a modified version of Lemma \ref{lemma2.1} for \(r\in[0,R]\) instead of \(r\in[0,\infty[\). In turn, the proof for case  \(r_1> r_c^-\) follows the steps of section 7 of  \cite{Costa-Mena}.

We now have the necessary new estimates to conclude the proof using the argument presented in \cite{Costa-Mena}. 
In fact, by Gr\"onwall's inequality we can now show that, by decreasing $x'$ if necessary, there exist constants $C_1,H_1>0$ such that 
\[
\mcal{E}_R(u_1)\leq C_1\tilde{x}_0e^{-H_1u_1}.
\]
 Then we choose \(\tilde{x}_0\) so that \(C_1\tilde{x}_0\leq x'/2\), to obtain \(\mcal{E}_R(u_1)\leq x'/2\).
This implies that \(\mcal{U}_R\) is open in \([0, U^*[\). By connectedness it follows that \(\mcal{U}_R=[0, U^*[\).
Since our last estimate is independent of $R$, we can show that
\[
\|\partial_rh(u,\cdot)\|_{L_r^{\infty,2-\delta}}\leq x'<\gamma,\ \forall u\in [0,U^*[\;,
\]
from which we can then conclude, using Theorem~\ref{thm3.2}, that $U^*=\infty$ and global existence follows. 
\end{proof}
%
\section{Improved decay and global properties of the solutions}
\label{decay-sec}
The goal of this section is to prove Theorem \ref{thm-decay}. We divide the proof in three steps where we estimate the solution $h$ of the integro-differential system along side with the solution $\phi$ of the original system and its derivatives up to second order.
\subsection{Preliminary decay estimates}
\label{decay-sol-sec}
We start by proving decay properties of solutions to the integro-differential problem \eqref{asdfg}:
\begin{lemma}
\label{thm4.4}
Under the assumptions of Theorem \ref{thm3.3}:
\begin{equation}
\|\partial_rh(u,\cdot)\|_{L_r^{\infty,2-\delta}}\lesssim e^{-(1+\delta/2)u} \label{eq48}
\end{equation}
and
\begin{equation}
|\partial_r^2h(u,r)|\lesssim e^{-(1+\delta/2)u},\  \text{ for }0\leq r\leq r_c^- \label{eq48_1},
\end{equation}
where $r_c^-$ was defined in \eqref{rc}.
Also, given \(R>0\), if \(\tilde{x}_0\leq \underline{x}(R)\), with \(\underline{x}(R)\) small enough, then there exist constants \(C_R,C_{r_c^-}>0\) such that
\begin{align}
\sup_{r\leq R}|\partial_rh(u,r)|&\leq C_Re^{-2u}, \text{ and }\label{eq49}\\
\sup_{r\leq r_c^-}|\partial_r^2h(u,r)|&\leq C_{r_c^-}e^{-2u}.\label{eq49_1}
\end{align}
Moreover the exists $\underline{h}(\infty)$ such that 
\begin{equation}
\label{12345}
\begin{split}
|h(u,r)-\underline{h}(\infty)|&\leq C_Re^{-2u},
\end{split}
\end{equation}
where $r\le R$.
\end{lemma}
\begin{proof}
The proof of \eqref{eq48} and \eqref{eq49} follows a similar strategy as in the proof of Theorem 3 of \cite{Costa-Mena} with small changes so we omit the details. 

Now let us establish estimate \eqref{eq48_1}. From \eqref{lemma3.1} we have
\begin{equation}
\label{partial2h-eq}
\begin{split}
\partial_r^2h(u_1,r_1)=&h_0''(r(0))e^{\int_0^{u_1}3G(s,r(s))ds}+2\int_0^{u_1}\partial_rG(u,r(u))\partial_rh(u,r(u))e^{\int_u^{u_1}3G(s,r(s))ds}du\\
&-\int_0^{u_1}\partial_rJ(u,r(u))\frac{h-\bar{h}}{r}(u,r(u))e^{\int_u^{u_1}3G(s,r(s))ds}du\\
&+\int_0^{u_1}\frac{J(u,r(u))}{r(u)}\left[2\frac{h-\bar{h}}{r}-\partial_rh\right](u,r(u))e^{\int_u^{u_1}3G(s,r(s))ds}du,
\end{split}
\end{equation}
and write
\[
|\partial_r^2h(u_1,r_1)|\lesssim I_1+2I_2+I_3+I_4,
\]
where
\begin{equation}
\label{I-eq}
\aligned
I_1&=|h_0''(r(0))e^{\int_0^{u_1}3Gds};\\
I_2&=\int_0^{u_1}|\partial_rG||\partial_rh|e^{\int_u^{u_1}3Gds}du\\
I_3&=\int_0^{u_1}|\partial_rJ|\frac{|h-\bar{h}|}{r}e^{\int_u^{u_1}3Gds}du\\
I_4&=\int_0^{u_1}\frac{|J|}{r}\left(\frac{|h-\bar{h}|}{r}+|\partial_rh|\right)e^{\int_u^{u_1}3Gds}du.
\endaligned
\end{equation}
From \eqref{exp-estimate} we obtain for small $r$
\[
\begin{split}
I_2&\lesssim \int_0^{u_1}\frac{e^{-(1+\delta/2)u}}{(1+r)^{2-\delta}}e^{3(1-\varepsilon_{x'})(u-u_1)}du\lesssim e^{-3u_1(1-\varepsilon_{x'})} \int_0^{u_1}e^{[-(1+\delta/2)+3(1-\varepsilon_{x'})]u}du \lesssim e^{-(1+\delta/2)u_1}.
\end{split}
\]
Similarly, we have
\[
\begin{split}
I_3&\lesssim \int_0^{u_1}\frac{\log^{\chi(\delta)}(e+r)}{(1+r)^{2-\delta^+}}e^{-(1+\delta/2)u}e^{\int_u^{u_1}3Gds}du\lesssim \int_0^{u_1}e^{-(1+\delta/2)u}e^{3(1-\varepsilon_{x'})(u-u_1)}du \lesssim e^{-(1+\delta/2)u_1}.
\end{split}
\]
These two estimates imply
\[
I_4\lesssim e^{-(1+\delta/2)u_1}.
\]
Finally,
\[
I_1\lesssim (1+r(0))^{3-\delta}|h_0''(r(0))|e^{-3(1-\varepsilon_{x'})u_1}\lesssim e^{-(1+\delta/2)u_1}.
\]
Hence, we see that estimate \eqref{eq48_1} holds. An analogous argument can be used to prove estimate \eqref{eq49_1}. 

Now for \(r\leq R\), Theorem \ref{thm3.3} implies that
\[
\begin{split}
|\partial_uh|&=|Dh+\frac{1}{2}\tilde{f}\partial_rh|\leq |G(h-\bar{h})|+\frac{1}{2}|\tilde{f}||\partial_rh|\\
&\lesssim (1+\varepsilon_{X_\delta})\frac{r^2}{2}\sup_{r\leq R}|\partial_rh(u,r)|+\frac{1}{2}|\tilde{f}|\sup_{r\leq R}|\partial_rh(u,r)|\lesssim C_Re^{-2u}.
\end{split}
\]
Since \(\exp(-2u)\) is integrable in \([0,+\infty[\), it follows that the limit of \(h(u,r)\), as \(u\rightarrow +\infty\), exists and is equal to
\[
h(0,r)+\int_0^{+\infty}\partial_uh(u,r)du=:\underline{h}(\infty,r),\text{ for }r\leq R.
\]
For \(r_1\leq r_2\leq R\) we have
\[
|h(u,r_1)-h(u,r_2)|\leq \int_{r_1}^{r_2}|\partial_rh(u,\rho)|d\rho\lesssim C_Re^{-2u}.
\]
Taking the limit as \(u\rightarrow +\infty\) we see that \(\underline{h}(\infty,r_1)=\underline{h}(\infty, r_2)\). So, actually \(\underline{h}(\infty,r)\equiv \underline{h}(\infty)\), for \(r\leq R\). Moreover,
\begin{equation}
\label{123}
\begin{split}
|h(u,r)-\underline{h}(\infty)|&\leq \int_u^{+\infty}|\partial_uh(s,r)|ds\leq C_R\int_u^{+\infty}e^{-2s}ds=C_Re^{-2u}
\end{split}
\end{equation}
and then \eqref{12345} follows.
\end{proof}
\begin{lemma}\label{lemma-asd}
There is some \(\gamma=\gamma(\delta)>0\) so that we have the estimates
\begin{eqnarray}
|\partial^2_{ur}h|\lesssim& \frac{1}{(1+r)^{1-\delta}}[\log(e+r)]^{3\chi(\delta)+1-H(\delta)}e^{-\gamma u}\\
\label{second-r-der}
|\partial_r^2h|\lesssim &\frac{1}{(1+r)^{3-\delta}}[\log(e+r)]^{3\chi(\delta)+1-H(\delta)}e^{-\gamma u}\\
\label{partial2-h}
|\partial^2_uh|\lesssim & (1+r)^{1+\delta}e^{-\gamma u}[\log(e+r)]^{4\chi(\delta)+1-H(\delta)}.
\end{eqnarray}
\end{lemma}
\begin{proof}
By the definition of \(D\) we have that \(\partial_uh=Dh+\tilde{f}\partial_rh/2=G(h-\bar{h})+\tilde{f}\partial_rh/2\). Therefore,
\[
\begin{split}
\partial_r\partial_uh&=\partial_r G(h-\bar{h})+2G\partial_rh-G\frac{h-\bar{h}}{r}+\frac{1}{2}\tilde{f}\partial_r^2h=2G\partial_rh-J\frac{h-\bar{h}}{r}+\frac{1}{2}\tilde{f}\partial_r^2h,
\end{split}
\]
where we used \(\partial_rG=(G-J)/r\). 

So far we have useful estimates for all the terms of the previous equation except for \(\partial^2_rh\). We may use \eqref{partial2h-eq} and \eqref{I-eq} and estimate separately  each term \(I_1, I_2, I_3\) and \(I_4\). From the proof of Theorem \ref{thm3.3} we have that
\[
e^{m\int_u^{u_1}G(s,r(s))ds}\lesssim e^{m(1-\varepsilon)(u-u_1)},
\]
for some \(\varepsilon>0\). Using this together with Proposition \ref{prop2.3}, we find uniform estimates in $r$ and $u$ as
\[
\aligned
I_1\lesssim &\frac{1}{(1+r(0))^{3-\delta}}e^{\int_0^{u_1}3qGds}e^{\int_0^{u_1}3(1-q)Gds}
\lesssim \frac{1}{(1+r(0))^{3-\delta}}\left(\frac{1+r(0)}{1+r_1}\right)^{6q-\eta}e^{-3(1-q)(1-\varepsilon)u_1}\\
\lesssim & \frac{e^{-3(1-q)(1-\varepsilon)u_1}}{(1+r_1)^{3-\delta}},
\endaligned
\]
choosing \(\eta=6q-3+\delta\) and \(q\in[0,1]\) large enough, for instance \(q>2/3\). For \(I_2\) we use Theorem \ref{thm3.3}, the proof of Proposition \ref{prop3.1} and Proposition \ref{prop2.3} to obtain
\[
\begin{split}
I_2&\lesssim \int_0^{u_1}\frac{e^{-(1+\delta/2)u}}{(1+r)^{2-\delta}}e^{\int_u^{u_1}3qGds}e^{\int_u^{u_1}3(1-q)Gds}du\lesssim \int_0^{u_1}\frac{e^{\int_u^{u_1}3qGds}}{(1+r)^{2-\delta}}e^{-(1+\delta/2)u}e^{3(1-q)(1-\varepsilon)(u-u_1)}du\\
&\lesssim e^{-3(1-q)(1-\varepsilon)u_1}\frac{1+u_1}{(1+r_1)^{3-\delta}},
\end{split}
\]
as long as \(-(1+\delta/2)+3(1-q)(1-\varepsilon)<0\) and \(2-\delta<6q-1\). But this can be achieved if \(q\) is sufficiently close to \(1\). Similarly, we can estimate \(I_3\) and \(I_4\) as:
\begin{align*}
I_3&\lesssim \int_0^{u_1}\frac{[\log(e+r)]^{3\chi+1-H}}{(1+r)^{4-3\delta^+}}e^{-(1+\delta/2)u}e^{\int_u^{u_1}3Gds}du\\
&\lesssim  \frac{[\log(e+r_1)]^{3\chi+1-H}(1+u_1)}{(1+r_1)^{5-3\delta^+}}e^{-3(1-q)(1-\varepsilon)u_1} \\
I_4&\lesssim \int_0^{u_1}\frac{[\log(e+r)]^{2\chi+1-H}}{(1+r)^{2-2\delta^+}}\left(\frac{\log^\chi(e+r)}{(1+r)^{2-\delta^+}}e^{-(1+\delta/2)u}+\frac{e^{-(1+\delta/2)u}}{(1+r)^{2-\delta}}\right)e^{\int_u^{u_1}3Gds}du\\
&\lesssim e^{-3(1-q)(1-\varepsilon)u_1}\left(\frac{[\log(e+r_1)]^{3\chi+1-H}}{(1+r_1)^{5-3\delta^+}}+\frac{[\log(e+r_1)]^{2\chi+1-H}}{(1+r_1)^{5-2\delta^+-\delta}}\right).
\end{align*}
which provide the estimate \eqref{second-r-der}.
Putting everything together we get
\begin{equation}
\label{mixed-der}
\aligned
|\partial_r\partial_uh|
&\lesssim \frac{[\log(e+r)]^{3\chi+1-H}(1+u)}{(1+r)^{1-\delta}}e^{-\gamma u},
\endaligned
\end{equation}
for \(\gamma>0\) small enough.
Finally, an expression for $\partial^2_uh$ is given by 
\[
\partial^2_uh=\frac{1}{2}\partial^2_{ur}\tilde{f}(h-\bar{h})+G(\partial_uh-\partial_u\bar{h})+\frac{\partial_u\tilde{f}}{2}\partial_rh+\frac{\tilde{f}}{2}\partial^2_{ur}h
\]
and we can estimate this as
\begin{equation*}
\begin{split}
|\partial^2_uh|
&\lesssim (1+r)^{1+\delta}(1+u)e^{-\gamma u}[\log(e+r)]^{4\chi(\delta)+1-H(\delta)}.
\end{split}
\end{equation*}
Note that we can absorb the factor $1+u$ by taking a smaller $\gamma$.
\end{proof}
\subsection{Decay of the matter field}
\begin{proof} [Proof of Theorem \ref{thm-decay}] The differentiability of $\phi$ follows from the coming Proposition \ref{diff}. We now prove the estimates \eqref{eq50}-\eqref{eq53}. To do that we use decay results for $h$ from the previous subsection \ref{decay-sol-sec}.

Assume $\phi_0\in C^{k+2}([0,+\infty[)$ as in Theorem \ref{thm-existence}.
Put \(h_0=\partial_r(r\phi_0)\). By our assumptions on \(\phi_0\) we know that \(h\in C^{k+1}([0,+\infty[)\), \(h_0\in L^\infty([0,+\infty[), h_0'\in L^{\infty,2-\delta}([0,+\infty[), h_0''\in L^{\infty, 3-\delta}([0,+\infty[)\). From Theorem \ref{thm3.3} we conclude that there exists some \(\tilde{x}_0\) such that, if \(\|h_0'\|_{L_r^{\infty, 2-\delta}}\leq \tilde{x}_0\), then the problem \eqref{asdfg} has a unique global solution \(h\in C^{k-1}([0,+\infty[\times [0,+\infty[)\). But then the function \(\phi:=\bar{h}\) solves the Einstein-scalar field system with the Bondi spherically symmetric metric \(g\) with associated functions \(f, \tilde{f}\). Note also that \(\phi=\bar{h}\in C^{k-1}([0,+\infty[\times [0,+\infty[)\). 

We now proceed to prove the estimates on \(\phi\). From \eqref{123} we get	
\begin{equation}
\label{phi-est}
|\phi(u,r)-\underline{h}(\infty)|\leq |h(u,r)-\bar{h}(u,r)|+|h(u,r)-\underline{h}(\infty)|\lesssim \frac{r}{2}\sup_{r\leq R}|\partial_rh(u,r)|+C_Re^{-2u}\leq C_Re^{-2u}.
\end{equation}
Noting that \(\partial_r\phi=(h-\bar{h})/r\), we get
\begin{equation}
\label{phi-est1}
|\partial_r\phi|\leq \frac{1}{2}\sup_{r\leq R}|\partial_rh(u,r)|\leq C_Re^{-2u}.
\end{equation}
Also, from 
\[
\partial^2_r\phi=\frac{1}{r}[\partial_rh-2\partial_r\phi]
\]
we see that
\begin{equation}
\label{phi-est2}
|\partial_r^2\phi|\leq C_Re^{-2u},\text{ for }r_c^-\leq r\leq R.
\end{equation}
To get such an estimate for small \(r\) we need to be more careful in computing \(\partial_r^2\phi\). We have that
\[
\begin{split}
\partial_r^2\phi(u,r)&=\frac{1}{r}[\partial_rh(u,r)-2\partial_r\phi(u,r)]=\frac{1}{r}[\partial_rh(u,r)-\frac{2}{r}(h(u,r)-\bar{h}(u,r))]\\
&=\frac{1}{r}\left[\partial_rh(u,r)-\frac{2}{r^2}\int_0^r(h(u,r)-h(u,s))ds\right]=\frac{1}{r}\left[\partial_rh(u,r)-\frac{2}{r^2}\int_0^r\int_s^r\partial_rh(u,\rho)d\rho ds\right]\\
&=\frac{2}{r^3}\int_0^r\int_s^r[\partial_rh(u,r)-\partial_rh(u,\rho)]d\rho ds=\frac{2}{r^3}\int_0^r\int_s^r\int_\rho^r\partial_r^2h(u,t)dtd\rho ds.
\end{split}
\]
A quick computation shows that $\frac{2}{r^3}\int_0^r\int_s^r\int_\rho^rdtd\rho ds=\frac{1}{3}$ and therefore
\[
|\partial_r^2\phi(u,r)|\lesssim \sup_{0\leq t\leq r}|\partial_r^2h(u,t)|.
\]
If \(r\leq r_c^-\) we can use \eqref{eq49_1} to obtain 
\begin{equation}
\label{phi-est3}
|\partial_r^2\phi|\lesssim C_{r_c^-}e^{-2u},
\end{equation}
which allows to shown estimate \eqref{eq53}, if we put \(\underline{\phi}(\infty)=\underline{h}(\infty)\). 

Now we turn to the proof of estimates \eqref{eq50}, \eqref{eq51} and \eqref{eq52}. We know from \eqref{eq48} that
\begin{equation*} 
 |\partial_rh(u,r)|\leq Ce^{-(1+\delta/2)u}(1+r)^{\delta-2},
 \end{equation*}
for $-1< \delta<1/2$. It follows that, for every \(u\geq 0\), \(\partial_rh(u,\cdot)\) is integrable in \(r\) from \(0\) to \(+\infty\). So, the limit of \(h(u,r)\) as \(r\rightarrow +\infty\) exists. We set
\[
\underline{\phi}(u)=\lim_{r\rightarrow +\infty}h(u,r).
\]
With this definition we have
\[
\begin{split}
|h(u,r)-\underline{\phi}(u)|&\leq \int_r^{+\infty}|\partial_rh(u,s)|ds\lesssim \int_r^{+\infty}\frac{1}{(1+s)^{2-\delta}}e^{-(1+\delta/2)u}ds\lesssim \frac{1}{(1+r)^{1-\delta}}e^{-(1+\delta/2)u}.
\end{split}
\]
Therefore we get
\[
\begin{split}
|\phi(u,r)-\underline{\phi}(u)|&=|\bar{h}(u,r)-\underline{\phi}(u)|\leq \frac{1}{r}\int_0^r|h(u,s)-\underline{\phi}(u)|ds\lesssim \frac{1}{r}\int_0^r\frac{1}{(1+s)^{1-\delta}}e^{-(1+\delta/2)u}ds\\
&\lesssim (\chi(\delta)+A(r)[1-\chi(\delta)])\frac{\log^{\chi(\delta)}(e+r)}{(1+r)^{1-\delta^+}}e^{-(1+\delta/2)u},
\end{split}
\]
where
\[
A(r)=\frac{(1+r)^{1-\delta^+}}{r}\left[\frac{(1+r)^\delta}{\delta}-\frac{1}{\delta}\right],
\]
which is only relevant when \(\delta\neq 0\).
It is easy to check that 
\[
\lim_{r\rightarrow 0}A(r)=1 \text{ and }\lim_{r\rightarrow +\infty}A(r)=\frac{1}{|\delta|},
\]
so
\[
|\phi(u,r)-\underline{\phi}(u)|\lesssim \frac{\log^{\chi(\delta)}(e+r)}{(1+r)^{1-\delta^+}}e^{-(1+\delta/2)u},
\]
which corresponds to estimate \eqref{eq50} of Theorem \ref{thm-decay}. 

We can now use this estimate to show that \(\underline{\phi}\) is continuous. Let \(u_n\rightarrow u\) and let \(\varepsilon>0\). For \(r\) large enough, say \(r>M\), we have
\[
|\underline{\phi}(u)-h(u,r)|\lesssim \frac{1}{(1+r)^{1-\delta^+}}<\frac{\varepsilon}{3},\ \forall u\geq 0.
\]
By continuity of \(h\) there is some \(N\) such that $n\geq N\implies |h(u_n,r)-h(u,r)|<\frac{\varepsilon}{3}$. Thus, for \(n\geq N\) and \(r>M\) we have
\[
|\underline{\phi}(u_n)-\underline{\phi}(u)|\leq |\underline{\phi}(u_n)-h(u_n,r)|+|h(u_n,r)-h(u,r)|+|h(u,r)-\underline{\phi}(u)|<\varepsilon.
\]
Hence \(\underline{\phi}\) is continuous. Also we have from Lemma \ref{lemma2.1} that 
\[
|\partial_r\phi|=|\partial_r\bar{h}|=\frac{|h-\bar{h}|}{r}\lesssim \frac{\log^{\chi(\delta)}(e+r)}{(1+r)^{2-\delta^+}}\|\partial_rh\|_{L_r^{\infty,2-\delta}}\lesssim \frac{\log^{\chi(\delta)}(e+r)}{(1+r)^{2-\delta^+}}e^{-(1+\delta/2)u},
\]
so we have shown estimate \eqref{eq51}. Taking one more derivative and assuming that \(r\geq r_c^-\) we obtain
\[
|\partial^2_r\phi|=\frac{1}{r}|\partial_rh-2\partial_r\phi|\lesssim \frac{\log^{\chi(\delta)}(e+r)}{(1+r)^{3-\delta^+}}e^{-(1+\delta/2)u}.
\]
For \(r\leq r_c^-\) we can use \eqref{eq48_1} to get
\[
|\partial_r^2\phi(u,r)|\lesssim \sup_{0\leq t\leq r}|\partial_r^2h(u,t)|\lesssim e^{-(1+\delta/2)u}.
\]
These two estimates taken together imply \eqref{eq52}. 

For the proof of \eqref{eq51a} we simply use $\partial_u\phi=\int_0^r  \partial_u h(u,s)ds$ and the estimate  \eqref{tfghb} for $\partial_u h$.
In turn, the proof of \eqref{eq52a} follows from the estimate for $|\partial_u^2 h|$ in Lemma \ref{lemma-asd}. 
Finally, \eqref{eq53} follows directly from the above estimates \eqref{phi-est}-\eqref{phi-est3} and this completes the proof of Theorem \ref{thm-decay}.
\end{proof}
\subsection{Properties of the asymptotic solutions}
In this subsection we prove some properties of the asymptotic solution $\underline{\phi}$ in particular its differentiability as part of Theorem \ref{thm-decay}.  We show:
\begin{prop}
\label{diff}
Let \(\delta\in ]-1,1/2[\). Then, \(\partial^2_{ur}h(u,\cdot)\in L^1(]0,+\infty[)\), \(\forall u\in[0,+\infty[\). Moreover, \(\underline{\phi}\in C^1(]0,+\infty[)\) and 
\[
\partial_u\underline{\phi}(u)=\partial_uh(u,0)+\int_0^{+\infty}\partial^2_{ur}h(u,\rho)d\rho. 
\]
\end{prop}
\begin{proof}
Let $-1<\delta<0$.
The fact that \(\partial^2_{ur}h(u,\cdot)\in L^1(]0,+\infty[)\) follows immediately from the estimate in Lemma \ref{lemma-asd}. From the definition of \(\underline{\phi}\) we have that 
\[
\frac{\underline{\phi}(u+t)-\underline{\phi}(u)}{t}=\frac{h(u+t,0)-h(u,0)}{t}+\int_0^{+\infty}\frac{1}{t}\int_u^{u+t}\partial^2_{ur}h(s,\rho)dsd\rho.
\]
Using the estimate from Lemma \ref{lemma-asd} we see that
\[
\left|\frac{1}{t}\int_u^{u+t}\partial^2_{ur}h(s,\rho)ds\right|\lesssim \frac{2+u}{(1+\rho)^{1-\delta/2}},
\]
when \(|t|\leq 1\). Since, for $\delta<0$, \((1+\cdot)^{\delta/2-1}\in L^1(]0,+\infty[)\), it follows by the dominated convergence theorem that
\[
\partial_u\underline{\phi}(u)=\partial_uh(u,0)+\int_0^{+\infty}\partial^2_{ur}h(u,\rho)d\rho. 
\]
Then, the continuity of \(\partial_u\underline{\phi}\) follows immediately from the fact that \(\partial^2_{ur}h\) is continuous and from Lemma \ref{lemma-asd}.

We can now extend the result to the remaining values of $\delta$ by  using the simple domain of dependence argument presented before Corollary~\ref{corIntro}.
\end{proof}
We end this subsection by proving two further properties of $\underline \phi$:
\begin{cor} As a consequence of Proposition \ref{diff}:
\begin{enumerate} 
\item 
$\lim_{u\rightarrow +\infty}\underline{\phi}(u)=\underline{\phi}(\infty).$
\item $\lim_{u\rightarrow +\infty}\partial_u\underline{\phi}(u)=0 \text{ when }\delta<0.$
\end{enumerate}
\end{cor}
\begin{proof}
Recall that for fixed \(r\),
\[
\underline{\phi}(\infty)=\lim_{u\rightarrow +\infty}h(u,r).
\]
In particular, \(\lim_{u\rightarrow +\infty}h(u,0)=\underline{\phi}(\infty)\). Now, note that
\[
\left|\int_0^{+\infty}\partial_rh(u,\rho)d\rho\right|\lesssim \int_0^{+\infty}\frac{e^{-(1+\delta/2)u}}{(1+\rho)^{2-\delta}}d\rho\lesssim e^{-(1+\delta/2)u}\xrightarrow[u\rightarrow +\infty]{}0.
\]
In this way we see that
\[
\lim_{u\rightarrow +\infty}\underline{\phi}(u)=\lim_{u\rightarrow +\infty}h(u,0)+\int_0^{+\infty}\partial_rh(u,\rho)d\rho=\underline{\phi}(\infty).
\]
Since \(\underline{\phi}\) has a limit at infinity, we expect that its derivative goes to zero. Indeed this is the case, at least when \(\delta<0\). In that case
\[
\left|\int_0^{+\infty}\partial^2_{ur}h(u,\rho)d\rho\right|\lesssim (1+u)e^{-\gamma u}\int_0^{+\infty}\frac{[\log(e+\rho)]^{3\chi(\delta)+1-H(\delta)}}{(1+\rho)^{1-\delta}}d\rho \xrightarrow[u\rightarrow +\infty]{}0.
\]
Also, from the equation \(\partial_uh=G(h-\bar{h})+\tilde{f}\partial_rh/2\), which comes directly from the integro-differential equation \eqref{eq3}, we can use Lemma \ref{lemma2.1}, Proposition \ref{prop1} and estimates \eqref{eq18} and \eqref{eq48} to obtain
\begin{equation}
\label{tfghb}
|\partial_uh(u,r)|\lesssim \frac{r^2\log^{\chi(\delta)}(e+r)}{(1+r)^{2-\delta^+}}e^{-(1+\delta/2)u}+(1+r)^2\frac{e^{-(1+\delta/2)u}}{(1+r)^{2-\delta}}.
\end{equation}
In particular,
\[
|\partial_uh(u,0)|\lesssim e^{-(1+\delta/2)u}\xrightarrow[u\rightarrow +\infty]{}0
\]
and thus the second property follows.
\end{proof}
\section{Asymptotic convergence to the de Sitter solution}
\label{converge-sec}
The objective of this section is to prove Theorem \ref{thm-stability}. We split the proof into three subsections corresponding to the convergence of the metric and its first and second order derivatives. 
\subsection{Convergence of the metric}
We start by showing the statement \eqref{eq54} of Theorem \ref{thm-stability} following the steps of \cite{Costa-Mena}.

Since \(f\) is bounded and increasing in \(r\), the limit of \(f(u,r)\) as \(r\rightarrow +\infty\) exists, for each \(u\geq 0\). Set
\[
f(u,\infty)=\lim_{r\rightarrow +\infty}f(u,r).
\]
Define a new coordinate \(\hat{u}\) by \(d\hat{u}=f(u,\infty)du\) or equivalently the function
\[
\hat{u}(u)=\int_0^uf(s,\infty)ds.
\]
Using
\[
\hat{u}(u)-\hat{u}(u_0)=\int_{u_0}^uf(s,\infty)ds
\]
and $1\leq f(u,r)\leq 1+\varepsilon_{X_\delta}$, we see that
\[
\hat{u}(u_0)-u_0+u\leq \hat{u}\leq (1+\varepsilon_{X_\delta})u+\hat{u}(u_0)-(1+\varepsilon_{X_\delta})u_0.
\]
Choosing \(u_0=0\) and \(\hat{u}(0)=0\) we get
\[
u\leq \hat{u}\leq (1+\varepsilon_{X_\delta})u.
\]
Since \(\hat{u}\) is strictly increasing it is invertible, so we often write \(u=u(\hat{u})\). In this way we define the functions
\[
f_1(\hat{u},r)=\frac{f(u(\hat{u}),r)}{f(u(\hat{u}),\infty)}\qquad\text{ and }\qquad f_2(\hat{u},r)=\frac{\tilde{f}(u(\hat{u}),r)}{f(u(\hat{u}),\infty)}.
\]
In these coordinates, the metric \eqref{metric}  becomes
\[
g=-f_1(\hat{u},r)f_2(\hat{u},r)d\hat{u}^2-2f_1(\hat{u},r)d\hat{u}dr+r^2\sigma_{\mathbb S^2}.
\]
Now we let 
\[
e_0=\frac{1}{\sqrt{r^2-1}}\partial_{\hat{u}}+\sqrt{r^2-1}\,\partial_r\qquad \text{ and }\qquad e_1=\frac{1}{\sqrt{r^2-1}}\,\partial_{\hat{u}},
\]
which together with some orthonormal basis \(e_2,e_3\) of \(\mathbb S^2\) forms an orthonormal basis of de Sitter spacetime. For this to be well defined we assume from now on that \(r\gg 1\). Now we write our metric in this basis. If we set \(g_{IJ}(\hat{u},r)=g(e_I,e_J)\) we get
\begin{align*}
    g_{00}(\hat{u},r)&=\frac{f_1(\hat{u},r)f_2(\hat{u},r)}{1-r^2}-2f_1(\hat{u},r)\\
    g_{01}(\hat{u},r)&=\frac{f_1(\hat{u},r)}{1-r^2}[f_2(\hat{u},r)-(1-r^2)]\\
    g_{11}(\hat{u},r)&=\frac{f_1(\hat{u},r)f_2(\hat{u},r)}{1-r^2}.
\end{align*}
Our goal is to prove convergence of \(\partial^{\beta_0}_{e_0}\partial^{\beta_1}_{e_1}g_{IJ}(\hat{u},r)\) to \(\partial_{e_0}^{\beta_0}\partial_{e_1}^{\beta_1}g_{IJ}^{dS}(\hat{u},r)\), where \(g_{00}^{dS}=-1, g_{01}^{dS}=0\), \(g_{11}^{dS}=1\) and \( \beta_0,\beta_1\geq 0\) with \(\beta_0+\beta_1\leq 2\). In this subsection we just deal with the metric itself whose convergence can simply be seen through
\begin{equation}
\label{abcde}
\begin{split}
|f(u,r)-f(u,\infty)|&=\left|\exp\left(\frac{1}{2}\int_0^r\frac{(h-\bar{h})^2}{s}ds\right)-\exp\left(\frac{1}{2}\int_0^\infty\frac{(h-\bar{h})^2}{s}ds\right)\right|\\
&\lesssim \int_r^\infty\frac{(h-\bar{h})^2}{s}ds\lesssim e^{-(2+\delta)u}\int_r^\infty\frac{\log^{2\chi(\delta)}(e+s)}{(1+s)^{3-2\delta^+}}ds\\
&\lesssim \frac{\log^{2\chi(\delta)}(e+r)}{(1+r)^{2-2\delta^+}}e^{-(2+\delta)(1-\varepsilon)\hat{u}},
\end{split}
\end{equation}
where we used Lemma  \ref{lemma2.1}, and
\[
\begin{split}
|\tilde{f}(u,r)-(1-r^2)f(u,\infty)|&=\frac{1}{r}\left|\int_0^r(1-3s^2)(f(u,s)-f(u,\infty))ds\right|\\
&\lesssim \frac{1}{r}e^{-(2+\delta)(1-\varepsilon)\hat{u}}\int_0^r(1+s)^2\frac{\log^{2\chi(\delta)}(e+s)}{(1+s)^{2-2\delta^+}}ds\\
&\lesssim \log^{2\chi(\delta)}(e+r)(1+r)^{2\delta^+}e^{-(2+\delta)(1-\varepsilon)\hat{u}}
\end{split}
\]
which can then be used to prove the statement \eqref{eq54} of Theorem \ref{thm-stability}, see also \cite{Costa-Mena}.
\subsection{Asymptotic $C^1$-stability}
%
In this section we prove estimates \eqref{estimates} of Theorem \ref{thm-stability}. To start with, we collect a number of estimates for the first derivatives of $f$ and $\tilde f$ that will be needed in the sequel:
\begin{lemma} Given \(\delta\in ]-1,1/2[\), there is some \(\gamma=\gamma(\delta)>0\) such that the following estimates hold:
\begin{equation}
\label{drtyg}
|(1-r^2)f-\tilde{f}|\lesssim [\log(e+r)]^{2\chi(\delta)+1-H(\delta)}e^{-(2+\delta)u}r^2(1+r)^{2\delta^+-2}
\end{equation}
\begin{eqnarray}
&|\partial_rf|\lesssim \displaystyle {\frac{r\log^{2\chi(\delta)}(e+r)}{(1+r)^{4-2\delta^+}}e^{-(2+\delta)u}\label{1st}}\\
&|\partial_r\tilde{f}+2r|\lesssim (1+r)e^{-(2+\delta)u}\label{2nd}\\
&|\partial_uf|\lesssim e^{-(1+\frac{\delta}{2}+\gamma)u}\label{3rd}\\
&|\partial_u\tilde{f}|\lesssim (1+r)^2e^{-(1+\frac{\delta}{2}+\gamma)u}\label{4th}
\end{eqnarray}
\end{lemma}
\begin{proof}
Estimate \eqref{1st} follows directly from Lemma \ref{lemma2.1}, equation \eqref{eq48} and
\[
\partial_rf=\frac{1}{2}\frac{(h-\bar{h})^2}{r}f.
\]
Now differentiating $f$ with respect to \(u\) we see that
\[
\partial_uf(u,r)=f(u,r)\int_0^r\frac{h(u,\rho)-\bar{h}(u,\rho)}{\rho}[\partial_uh(u,\rho)-\partial_u\bar{h}(u,\rho)]d\rho. 
\]
We can estimate \(|\partial_uh-\partial_u\bar{h}|\) using
\[
\frac{1}{r}\int_0^r\int_\rho^r|\partial^2_{ru}h(u,s)|dsd\rho\lesssim [\log(e+r)]^{3\chi(\delta)+1-H(\delta)}\frac{(1+u)e^{-\gamma u}}{r}\int_0^r\frac{r-\rho}{(1+\rho)^{1-\delta}}d\rho,
\]
which comes from \eqref{mixed-der}. If \(\delta=0\) then the last integral is just equal to \(r\log(1+r)-r+\log(1+r)\) which can be estimated by \(r^2(1+r)^{-1}\log(e+r)\). If \(\delta\neq 0\), then the integral equals
\[
\frac{(1+r)^{\delta+1}}{\delta(\delta+1)}-\frac{1}{\delta(\delta+1)}-\frac{r}{\delta},
\]
which is estimated by \(r^2(1+r)^{\delta-1}\). Therefore we get
\begin{equation}
\label{diff-in-h}
|\partial_uh(u,r)-\partial_u\bar{h}(u,r)|\lesssim [\log(e+r)]^{4\chi(\delta)+1-H(\delta)}(1+u)e^{-\gamma u}r(1+r)^{\delta-1}.
\end{equation}
We can now use this to obtain an estimate for \(\partial_uf\) and show \eqref{3rd} as
\[
|\partial_uf|\lesssim (1+u)e^{-(1+\frac{\delta}{2}+\gamma)u}\int_0^r\frac{[\log(e+\rho)]^{5\chi(\delta)+1-H(\delta)}}{(1+\rho)^{2-\delta^+-\delta}}d\rho\lesssim (1+u)e^{-(1+\frac{\delta}{2}+\gamma)u}.
\]
Recalling that 
\[
\tilde{f}(u,r)=\frac{1}{r}\int_0^r(1-3\rho^2)f(u,\rho)d\rho, 
\]
we find
\[
\begin{split}
|\partial_u\tilde{f}(u,r)|&\leq \frac{1}{r}\int_0^r|1-3\rho^2||\partial_uf(u,\rho)|d\rho \lesssim (1+r)^2(1+u)e^{-(1+\frac{\delta}{2}+\gamma)u}.
\end{split}
\]
Finally, we estimate $|\partial_r\tilde{f}+2r|$ using
\[
|\partial_r\tilde{f}+2r|\leq \frac{1}{r}|(1-r^2)f-\tilde{f}|+2r|f-1|.
\]
From
\[
(1-r^2)f-\tilde{f}=\frac{1}{r}\int_0^r(1-3\rho^2)\int_\rho^r\partial_rf(u,y)dyd\rho,
\]
we see that
\[
\begin{split}
|(1-r^2)f-\tilde{f}|&\lesssim \frac{1}{r}\int_0^r|1-3\rho^2|\int_\rho^r\frac{y\log^{2\chi(\delta)}(e+y)}{(1+y)^{4-2\delta^+}}e^{-(2+\delta)u}dyd\rho \\
&\lesssim \log^{2\chi(\delta)}(e+r)e^{-(2+\delta)u}\frac{1}{r}\int_0^r|1-3\rho^2|\frac{r-\rho}{(1+\rho)^{3-2\delta^+}}d\rho\\
&\lesssim [\log(e+r)]^{2\chi(\delta)+1-H(\delta)}e^{-(2+\delta)u}r^2(1+r)^{2\delta^+-2}
\end{split}
\]
and therefore
\[
|\partial_r\tilde{f}+2r|\lesssim [\log(e+r)]^{2\chi(\delta)+1-H(\delta)}e^{-(2+\delta)u}(1+r)^{2\delta^+-1}+re^{-(2+\delta)u}\lesssim (1+r)e^{-(2+\delta)u}.
\]
\end{proof}
We now have the tools to estimate \(\partial_rg_{01}\). First observe that 
\[
\partial_rg_{01}=\frac{f_1}{1-r^2}[\partial_rf_2+2r]+[f_2-(1-r^2)]\frac{\partial_rf_1(1-r^2)+2rf_1}{(1-r^2)^2}.
\]
We can estimate \(\partial_rf_2+2r\) using \eqref{drtyg} and \eqref{abcde} as
\[
\begin{split}
  |\partial_rf_2+2r|&=\frac{1}{f(u,\infty)}|\partial_r\tilde{f}+2rf(u,\infty)|\lesssim\frac{1}{r}|(1-r^2)f-\tilde{f}|+r|f(u,r)-f(u,\infty)|  \\
  &\lesssim \frac{[\log(e+r)]^{2\chi(\delta)+1-H(\delta)}}{(1+r)^{1-2\delta^+}}e^{-(2+\delta)(1-\varepsilon)\hat{u}}.
\end{split}
\]
Furthermore, using \eqref{1st} we have that
\[
|\partial_rf_1(1-r^2)+2rf_1|=\frac{1}{f(u,\infty)}|\partial_rf(1-r^2)+2rf|\lesssim r.
\]
Therefore, putting these together we get
\begin{equation}
\label{pag32}
\begin{split}
|\partial_rg_{01}|
&\lesssim \frac{[\log(e+r)]^{2\chi(\delta)+1-H(\delta)}}{(1+r)^{3-2\delta^+}}e^{-(2+\delta)(1-\varepsilon)\hat{u}}.
\end{split}
\end{equation}
Now we turn our attention to \(\partial_{\hat{u}}g_{01}\). We have that
\begin{equation}
\label{pag32-2}
\partial_{\hat{u}}g_{01}=\frac{\partial_{\hat{u}}f_1}{1-r^2}[f_2-(1-r^2)]+\frac{f_1}{1-r^2}\partial_{\hat{u}}f_2.
\end{equation}
To obtain an estimate for this derivative we need to control \(\partial_{\hat{u}}f_1\) and \(\partial_{\hat{u}}f_2\). We have that 
\[
\partial_{\hat{u}}f_1=\frac{\partial_uf-\hat{f}\partial_uf(u,\infty)}{f(u,\infty)^2}=\frac{\partial_uf-\partial_uf(u,\infty)-[\hat{f}-1]\partial_uf(u,\infty)}{f(u,\infty)^2}.
\]
First we estimate
\[
\begin{split}
  |\partial_uf(u,r)-\partial_uf(u,\infty)|&\leq |f(u,r)-f(u,\infty)|\int_0^r\frac{|h-\bar{h}|}{\rho}|\partial_uh-\partial_u\bar{h}|d\rho\\
  &+f(u,\infty)\int_r^\infty \frac{|h-\bar{h}|}{\rho}|\partial_uh-\partial_u\bar{h}|d\rho\\
  &\lesssim (1+\hat{u})\frac{[\log(e+r)]^{5\chi(\delta)+1-H(\delta)}}{(1+r)^{1-\delta^+-\delta}}e^{-(1+\delta/2+\gamma)(1-\varepsilon)\hat{u}},
\end{split}
\]
where we used \eqref{diff-in-h}. So we have that
\[
\begin{split}
|\partial_{\hat{u}}f_1|
&\lesssim (1+\hat{u})\frac{[\log(e+r)]^{5\chi(\delta)+1-H(\delta)}}{(1+r)^{1-\delta^+-\delta}}e^{-(1+\delta/2+\gamma)(1-\varepsilon)\hat{u}}.
\end{split}
\]
As for \(\partial_{\hat{u}}f_2\) we may write
\begin{equation}
\label{eq23456}
\partial_{\hat{u}}f_2=\frac{\partial_u\tilde{f}-\tilde{f}\partial_uf(u,\infty)/f(u,\infty)}{f(u,\infty)^2}.
\end{equation}
We can take advantage of some cancellation here by writing
\begin{equation}
\label{eq09876}
\aligned
\partial_u\tilde{f}-\tilde{f}\frac{\partial_uf(u,\infty)}{f(u,\infty)}&=\frac{1}{r}\int_0^r(1-3s^2)\partial_uf(u,s)ds-\frac{1}{r}\int_0^r(1-3s^2)f(u,s)ds\int_0^\infty \frac{h-\bar{h}}{\rho}(\partial_uh-\partial_u\bar{h})d\rho \\
&=-\frac{1}{r}\int_0^r(1-3s^2)f(u,s)\int_s^\infty\frac{h-\bar{h}}{\rho}(\partial_uh-\partial_u\bar{h})d\rho ds. 
\endaligned
\end{equation}
We can now estimate this using \eqref{diff-in-h} as
\begin{equation*}
\begin{split}
  |\partial_{\hat{u}}f_2|&\lesssim   (1+\hat{u})e^{-(1+\delta/2+\gamma)(1-\varepsilon)\hat{u}}[\log(e+r)]^{5\chi(\delta)+1-H(\delta)}(1+r)^{1+\delta^++\delta}.
\end{split}
\end{equation*}
With everything together we see that from \eqref{pag32-2}
\begin{equation}
\label{pag-33}
|\partial_{\hat{u}}g_{01}|\lesssim \frac{[\log(e+r)]^{7\chi(\delta)+1-H(\delta)}}{(1+r)^{1-\delta^+-\delta}}(1+\hat{u})e^{-(1+\delta/2+\gamma)(1-\varepsilon)\hat{u}}.
\end{equation}
Now that we have the first order derivatives \eqref{pag32} and \eqref{pag-33} for \(g_{01}\) we can compute the first order directional derivatives. We then get
\[
|\partial_{e_0}g_{01}|\lesssim \frac{[\log(e+r)]^{7\chi(\delta)+1-H(\delta)}}{(1+r)^{2-2\delta^+}}(1+\hat{u})e^{-(1+\delta/2+\gamma)(1-\varepsilon)\hat{u}}
\]
and
\[
|\partial_{e_1}g_{01}|\lesssim \frac{[\log(e+r)]^{7\chi(\delta)+1-H(\delta)}}{(1+r)^{2-\delta^+-\delta}}(1+\hat{u})e^{-(1+\delta/2+\gamma)(1-\varepsilon)\hat{u}},
\]
where we can absorb the factor $1+\hat u$ by taking a smaller $\gamma$.

Estimates for the first order directional derivatives of the remaining metric coefficients \(g_{00}\) and \(g_{11}\) follow by an identical procedure since we can write \(g_{00}=g_{01}-f_1$ and $g_{11}=g_{01}+f_1\). In fact this yields identical estimates as for \(g_{01}\) which then proves estimates \eqref{estimates} of Theorem \ref{thm-stability}.
\subsection{Asymptotic $C^2$-stability}
We start by showing the following estimates for the second order derivatives of $f$:
\begin{lemma}
Given \(\delta\in ]-1,1/2[\), there is some \(\gamma=\gamma(\delta)>0\) such that the following estimates hold:
\begin{align*}
&|\partial^2_rf|\lesssim \frac{\log^{4\chi(\delta)}}{(1+r)^{4-2\delta^+}}e^{-(2+\delta)u}\\
    &|\partial^2_{ur}f|\lesssim \frac{[\log(e+r)]^{5\chi(\delta)+1-H(\delta)}}{(1+r)^{2-\delta^+-\delta}}e^{-(1+\frac{\delta}{2}+\gamma)u}\\
    &|\partial^2_uf|\lesssim \log^{11\chi(\delta)}(e+r)(1+r)^{2\delta^+}e^{-2\gamma u}
\end{align*}
\end{lemma}
\begin{proof}
For the second derivative with respect to \(r\), we have the expression
\[
\partial_r^2f=f\frac{h-\bar{h}}{r}\partial_rh-\frac{3}{2}f\frac{(h-\bar{h})^2}{r^2}+\frac{1}{4}\frac{(h-\bar{h})^4}{r^2}f.
\]
We can estimate every term of this expression using Lemma \ref{lemma2.1} to obtain
\[
|\partial^2_rf|\lesssim \frac{\log^{4\chi(\delta)}(e+r)}{(1+r)^{4-2\delta^+}}e^{-(2+\delta)u}.
\]
We now turn to the derivatives that involve \(u\). We start with
\[
\partial^2_{ur}f=\partial_rf\int_0^r\frac{h-\bar{h}}{\rho}[\partial_uh-\partial_u\bar{h}]d\rho+f\frac{h-\bar{h}}{r}[\partial_uh-\partial_u\bar{h}].
\]
Therefore using Lemma \ref{lemma2.1} and \eqref{diff-in-h}  we get
\[
\begin{split}
|\partial^2_{ur}f|
&\lesssim \frac{[\log(e+r)]^{5\chi(\delta)+1-H(\delta)}}{(1+r)^{2-\delta^+-\delta}}(1+u)e^{-(1+\frac{\delta}{2}+\gamma)u}.
\end{split}
\]
Now we can estimate \(\partial_u^2f\) using
\[
\partial_u^2f=\partial_uf\int_0^r\frac{h-\bar{h}}{\rho}[\partial_uh-\partial_u\bar{h}]d\rho+f\int_0^r\frac{(\partial_uh-\partial_u\bar{h})^2}{\rho}d\rho+f\int_0^r\frac{h-\bar{h}}{\rho}[\partial_u^2h-\partial_u^2\bar{h}]d\rho. 
\]
and the estimate \eqref{partial2-h} for \(\partial^2_uh\). In fact since the estimate for \(\partial^2_u\bar{h}\) is the same as for \(\partial^2_uh\) we get
\[
\begin{split}
  |\partial_u^2f|
  &\lesssim (1+u)^2e^{-2\gamma u}[\log(e+r)]^{11\chi(\delta)}(1+r)^{2\delta^+}.
\end{split}
\]
\end{proof}
We consider now \(\partial_{\hat{u}}^2g_{01}\). First observe that
\[
\partial_{\hat{u}}^2g_{01}=\frac{\partial_{\hat{u}}^2f_1}{1-r^2}[f_2-(1-r^2)]+2\frac{\partial_{\hat{u}}f_1\partial_{\hat{u}}f_2}{1-r^2}+\frac{f_1}{1-r^2}\partial^2_{\hat{u}}f_2.
\]
To estimate this we must be able to control \(\partial_{\hat{u}}^2f_1\) and \(\partial_{\hat{u}}^2f_2\). We start by noting that
\begin{equation}
\label{pag34}
\begin{split}
\partial^2_{\hat{u}}f_1&=\frac{1}{f(u,\infty)}\left(\partial_u^2f-\hat{f}\partial^2_uf(u,\infty)-\partial_uf(u,\infty)\frac{\partial_uf-\hat{f}\partial_uf(u,\infty)}{f(u,\infty)}\right)\\
&-2\frac{\partial_uf(u,\infty)}{f(u,\infty)^2}(\partial_uf-\hat{f}\partial_uf(u,\infty))
\end{split}
\end{equation}
and we focus on estimating \(\partial^2_uf-\hat{f}\partial_u^2f(u,\infty)\). Here we have to be careful. We want to say that
\[
\begin{split}
\partial^2_uf(u,\infty)&=f(u,\infty)\left(\int_0^\infty\frac{h-\bar{h}}{\rho}(\partial_uh-\partial_u\bar{h})d\rho\right)^2+f(u,\infty)\int_0^\infty\frac{(\partial_uh-\partial_u\bar{h})^2}{\rho}d\rho\\
&+f(u,\infty)\int_0^\infty\frac{h-\bar{h}}{\rho}(\partial^2_uh-\partial^2_u\bar{h})d\rho,
\end{split}
\]
however, given our estimates we can only guarantee that these integrals converge if \(\delta<0\). For this reason, when estimating second derivatives we assume that \(\delta<0\). We can then write
\[
\begin{split}
\partial^2_uf-\hat{f}\partial_u^2f(u,\infty)&=f(u,r)\left[\left(\int_0^r\frac{h-\bar{h}}{\rho}(\partial_uh-\partial_u\bar{h})d\rho\right)^2-\left(\int_0^\infty\frac{h-\bar{h}}{\rho}(\partial_uh-\partial_u\bar{h})d\rho\right)^2\right]\\
&-f\int_r^\infty\frac{(\partial_uh-\partial_u\bar{h})^2}{\rho}d\rho-f\int_r^\infty\frac{h-\bar{h}}{\rho}(\partial_u^2h-\partial^2_u\bar{h})d\rho
\end{split}
\]
which, from \eqref{diff-in-h}, \eqref{partial2-h} and Lemma \ref{lemma2.1}, results in
\[
\begin{split}
    |\partial_u^2f-\hat{f}\partial^2_uf(u,\infty)|
    &\lesssim [\log(e+r)]^{8\chi(\delta)+2-2H(\delta)}(1+r)^\delta (1+\hat{u})^2e^{-2\gamma(1-\varepsilon)\hat{u}}.
\end{split}
\]
Using this result we now estimate \(\partial^2_{\hat{u}}f_1\) from \eqref{pag34} as
\[
\begin{split}
    |\partial^2_{\hat{u}}f_1|
    &\lesssim [\log(e+r)]^{8\chi(\delta)+2-2H(\delta)}(1+r)^\delta (1+\hat{u})^2e^{-2\gamma(1-\varepsilon)\hat{u}}.
\end{split}
\]
Regarding \(\partial^2_{\hat{u}}f_2\) observe that
\[
\partial_{\hat{u}}^2f_2=\frac{1}{f(u,\infty)^3}\left(\partial^2_u\tilde{f}-\tilde{f}\frac{\partial_u^2f(u,\infty)}{f(u,\infty)}\right)-3\frac{\partial_uf(u,\infty)}{f(u,\infty)^4}\left(\partial_u\tilde{f}-\tilde{f}\frac{\partial_uf(u,\infty)}{f(u,\infty)}\right). 
\]
We know how to estimate the second term, so we focus on the first term. We can write
\[
\partial^2_u\tilde{f}-\tilde{f}\frac{\partial^2_uf(u,\infty)}{f(u,\infty)}=\frac{1}{r}\int_0^r(1-3s^2)(\partial^2_uf(u,s)-\hat{f}(u,s)\partial^2_uf(u,\infty))ds
\]
and therefore
\[
\begin{split}
\left|\partial^2_u\tilde{f}-\tilde{f}\frac{\partial^2_uf(u,\infty)}{f(u,\infty)}\right|
&\lesssim [\log(e+r)]^{8\chi(\delta)+2-2H(\delta)}(1+r)^{2+\delta}(1+\hat{u})^2e^{-2\gamma(1-\varepsilon)\hat{u}}.
\end{split}
\]
Using \eqref{eq09876}, this leads to the estimate
\[
|\partial^2_{\hat{u}}f_2|\lesssim [\log(e+r)]^{8\chi(\delta)+2-2H(\delta)}(1+r)^{2+\delta}(1+\hat{u})^2e^{-2\gamma(1-\varepsilon)\hat{u}}.
\]
Finally, we can put these terms together to obtain
\[
|\partial^2_{\hat{u}}g_{01}|\lesssim \log^2(e+r)(1+r)^\delta(1+\hat{u})^2e^{-2\gamma(1-\varepsilon)\hat{u}}.
\]
With this estimate we can now obtain the corresponding estimate for the second order derivative with respect to \(e_1\) as
\[
|\partial^2_{e_1}g_{01}|\lesssim \frac{\log^2(e+r)}{(1+r)^{2-\delta}}(1+\hat{u})^2e^{-2\gamma(1-\varepsilon)\hat{u}}.
\]
Again, this estimate works only for large \(r\) and \(\delta<0\). Next we estimate \(\partial^2_rg_{01}\). We have that
\[
\begin{split}
\partial_r^2g_{01}&=2(\partial_rf_2+2r)\left(\frac{\partial_rf_1}{1-r^2}+\frac{2r}{(1-r^2)^2}f_1\right)+\frac{f_1}{1-r^2}(\partial_r^2f_2+2)\\
&+(f_2-(1-r^2))\left(\frac{\partial^2_rf_1}{1-r^2}+\frac{2f_1}{(1-r^2)^2}+\frac{4r\partial_rf_1}{(1-r^2)^2}+\frac{8r^2f_1}{(1-r^2)^3}\right).
\end{split}
\]
So we are only missing the estimate for \(\partial^2_rf_2+2\) which can be obtained as
\[
\begin{split}
    |\partial^2_rf_2+2|&\lesssim (1+r)|\partial_rf|+\frac{1}{r^2}|(1-r^2)f-\tilde{f}|+|f(u,r)-f(u,\infty)|\\
    &\lesssim \frac{[\log(e+r)]^{2\chi(\delta)+1-H(\delta)}}{(1+r)^{2-2\delta^+}}e^{-(2+\delta)(1-\varepsilon)\hat{u}}.
\end{split}
\]
Using this expression we find
\[
|\partial^2_rg_{01}|\lesssim \frac{[\log(e+r)]^{6\chi(\delta)+1-H(\delta)}}{(1+r)^{4-2\delta^+}}e^{-(2+\delta)(1-\varepsilon)\hat{u}}.
\]
We now turn our attention to \(\partial^2_{r\hat{u}}g_{01}\). We may write
\begin{equation}
\label{pag36}
\begin{split}
\partial_{r\hat{u}}^2g_{01}&=(f_2-(1-r^2))\left(\frac{\partial^2_{r\hat{u}}f_1}{1-r^2}+\frac{2r\partial_{\hat{u}}f_1}{(1-r^2)^2}\right)+\frac{\partial_{\hat{u}}f_1}{1-r^2}(\partial_rf_2+2r)\\
&+\partial_{\hat{u}}f_2\left(\frac{\partial_rf_1}{1-r^2}+\frac{2rf_1}{(1-r^2)^2}\right)+\frac{f_1}{1-r^2}\partial^2_{r\hat{u}}f_2
\end{split}
\end{equation}
and we have estimates for all terms except for \(\partial^2_{r\hat{u}}f_1\) and \(\partial^2_{r\hat{u}}f_2\). Let's consider first \(\partial^2_{r\hat{u}}f_1\) written as
\[
\partial^2_{r\hat{u}}f_1=\frac{1}{f(u,\infty)^2}\left(\partial^2_{ur}f-\frac{\partial_uf(u,\infty)}{f(u,\infty)}\partial_rf\right).
\]
So,
\[
|\partial^2_{r\hat{u}}f_1|\lesssim \frac{[\log(e+r)]^{5\chi(\delta)+1-H(\delta)}}{(1+r)^{2-\delta^+-\delta}}(1+\hat{u})e^{-(1+\delta/2+\gamma)(1-\varepsilon)\hat{u}}.
\]
Next we estimate
\[
\begin{split}
\partial^2_{r\hat{u}}f_2&=\frac{1}{f(u,\infty)^2}\left(\partial^2_{ur}\tilde{f}-\partial_r\tilde{f}\frac{\partial_uf(u,\infty)}{f(u,\infty)}\right)\\
&=\frac{1}{f(u,\infty)^2}\left(-\frac{1}{r}\left(\partial_u\tilde{f}-\tilde{f}\frac{\partial_uf(u,\infty)}{f(u,\infty)}\right)-\frac{1}{r}(1-3r^2)f\int_r^\infty\frac{h-\bar{h}}{\rho}(\partial_uh-\partial_u\bar{h})d\rho\right)
\end{split}
\]
giving
\[
|\partial^2_{r\hat{u}}f_2|\lesssim [\log(e+r)]^{5\chi(\delta)+1-H(\delta)}(1+r)^{\delta^++\delta}(1+\hat{u})e^{-(1+\delta/2+\gamma)(1-\varepsilon)\hat{u}}.
\]
Finally, with these estimates we obtain from \eqref{pag36}
\[
|\partial^2_{r\hat{u}}g_{01}|\lesssim \frac{[\log(e+r)]^{7\chi(\delta)+2-2H(\delta)}}{(1+r)^{2-3\delta^+-\delta}}(1+\hat{u})e^{-(1+\delta/2+\gamma)(1-\varepsilon)\hat{u}}.
\]
Now that we have all the estimates for the derivatives of \(g_{01}\) up to second order we may estimate the remaining directional derivatives. We get
\[
|\partial^2_{e_0}g_{01}|\lesssim \frac{\log^2(e+r)}{(1+r)^2}(1+\hat{u})^2e^{-2\gamma(1-\varepsilon)\hat{u}}
\]
and
\[
|\partial_{e_0}\partial_{e_1}g_{01}|\lesssim \frac{\log^2(e+r)}{(1+r)^{2-\delta}}(1+\hat{u})^2e^{-2\gamma(1-\varepsilon)\hat{u}}.
\]
where for both cases we need \(\delta<0\). 

Finally, we obtain estimates for the derivatives of \(g_{00}\) and \(g_{11}\) by writting \(g_{00}=g_{01}-f_1,\ g_{11}=g_{01}+f_1\) and using the above estimates for the derivatives of \(f_1\). Doing this, yields the same estimates as for \(g_{01}\) and therefore we obtain \eqref{estimates2} which concludes the proof of Theorem \ref{thm-stability}.

As a final note we remark that some of the above decay estimates can easily be improved for fixed $r$:
\begin{remark}
\label{cor-B}
\label{pointwise-decay}
For fixed $r>R$ we have
for any multi-index \(\beta=(\beta_u,\beta_r)\in \N_0^2\) with \(|\beta|\leq 2\) and $\partial^\beta=\partial^{\beta_u} \partial^{\beta_r}$, the following estimates hold for the metric and its derivatives:
\begin{equation}
\label{qwert}
\aligned
\|\partial^\beta f(u,\cdot)-\partial^\beta f^{dS}(u,\cdot)\|_{L^\infty([0,R])}&\lesssim_R(1+u)^{\chi(\beta_u-2)}e^{-[4-(2-\gamma)\chi(\beta_u-2)]u}\\
\|\partial^\beta \tilde{f}(u,\cdot)-\partial^\beta \tilde{f}^{dS}(u,\cdot)\|_{L^\infty([0,R])}&\lesssim_R(1+u)^{\chi(\beta_u-2)}e^{-[4-(2-\gamma)\chi(\beta_u-2)]u},
\endaligned
\end{equation}
where \(f^{dS}(u,r)=1\), \(\tilde{f}^{dS}(u,r)=1-r^2\) and \(\gamma\) is a positive constant.
\end{remark}
\section*{Acknowledgments}
The authors thank FCT/Portugal through CAMGSD, IST-ID, projects UIDB/04459/2020 and UIDP/04459/2020. This work was also partially supported by FCT/Portugal and CERN through the project CERN/FIS-PAR/0023/2019. RLD thanks FCT for the PhD grant PD/BD/150338/2019.

\small 

\appendix 

\section{The sequence \((h_n)_n\) contracts}
\label{appendix-A}
The proof of Theorem \ref{thm3.2} is based on showing that the sequence \((h_n)_n\) contracts in \(L_U^\infty L_r^\infty\) as we show next.

\begin{prop}\label{prop3.2}
Let \(-1\leq \delta<1/2\). For \(\|h_0'\|_{L_U^\infty L_r^{\infty, 2-\delta}}\) and \(U\) sufficiently small, the sequence \((h_n)_n\) contracts in \(L_U^\infty L_r^\infty\).
\end{prop}
\begin{proof}
Using \eqref{eq29},
\[
D_{n-1}w_n=2G_{n-1}w_n-J_{n-1}\frac{h_{n-1}-\bar{h}_{n-1}}{r}.
\]
But also, by definition,
\[
D_{n-1}w_n=\partial_uw_n-\frac{\tilde{f}_{n-1}}{2}\partial_rw_n.
\]
So we get
\[
\partial_uw_n=2G_{n-1}w_n-J_{n-1}\frac{h_{n-1}-\bar{h}_{n-1}}{r}+\frac{\tilde{f}_{n-1}}{2}\partial_rw_n.
\]
From this we see that
\[
\begin{split}
D_n(w_{n+1}-w_n)&=2G_nw_{n+1}-J_n\frac{h_n-\bar{h}_n}{r}+\frac{1}{2}\tilde{f}_n\partial_rw_n-\partial_uw_n\\
&=2G_nw_{n+1}-J_n\frac{h_n-\bar{h}_n}{r}+\frac{1}{2}\tilde{f}_n\partial_rw_n-2G_{n-1}w_n+J_{n-1}\frac{h_{n-1}-\bar{h}_{n-1}}{r}-\frac{1}{2}\tilde{f}_{n-1}\partial_rw_n\\
&=2G_n(w_{n+1}-w_n)+2(G_n-G_{n-1})w_n+\frac{\tilde{f}_n-\tilde{f}_{n-1}}{2}\partial_rw_n-\frac{J_n}{r}[(h_n-\bar{h}_n)-(h_{n-1}-\bar{h}_{n-1})]\\
&-\frac{J_n-J_{n-1}}{r}[h_{n-1}-\bar{h}_{n-1}].
\end{split}
\]
Again, if we consider this equation along the characteristics \((u,r_n(u;u_1,r_1))\), multiply it by an integrating factor and simplify the expression we are led to
\[
\begin{split}
\partial_u\left([w_{n+1}-w_n]e^{\int_u^{u_1}2G_nds}\right)&=2(G_n-G_{n-1})w_ne^{\int_u^{u_1}2G_nds}+\frac{\tilde{f}_n-\tilde{f}_{n-1}}{2}\partial_rw_ne^{\int_u^{u_1}2G_nds}\\
&-\frac{J_n}{r_n}[(h_n-\bar{h}_n)-(h_{n-1}-\bar{h}_{n-1})]e^{\int_u^{u_1}2G_nds}-\frac{J_n-J_{n-1}}{r_n}[h_{n-1}-\bar{h}_{n-1}]e^{\int_u^{u_1}2G_nds}.
\end{split}
\]
If we now integrate this in \(u\) from \(0\) to \(u_1\), and noting that \(w_{n+1}(0,r_n(0))-w_n(0,r_n(0))=h_0'(r_n(0))-h_0'(r_n(0))=0\), we get
\[
\begin{split}
w_{n+1}(u_1,r_1)-w_n(u_1,r_1)&=\int_0^{u_1}2(G_n-G_{n-1})w_ne^{\int_u^{u_1}2G_nds}du+\int_0^{u_1}\frac{\tilde{f}_n-\tilde{f}_{n-1}}{2}\partial_rw_ne^{\int_u^{u_1}2G_nds}du\\
&-\int_0^{u_1}\frac{J_n}{r_n}[(h_n-\bar{h}_n)-(h_{n-1}-\bar{h}_{n-1})]e^{\int_u^{u_1}2G_nds}du\\
&-\int_0^{u_1}\frac{J_n-J_{n-1}}{r_n}[h_{n-1}-\bar{h}_{n-1}]e^{\int_u^{u_1}2G_nds}du.
\end{split}
\]
This way we get the estimate
\[
|w_{n+1}(u_1,r_1)-w_n(u_1,r_1)|\leq I_1+I_2+I_3+I_4,
\]
where
\begin{align}
&I_1=2\int_0^{u_1}|G_n-G_{n-1}||w_n|e^{\int_u^{u_1}2G_nds}du;\vspace{10pt}\\
&I_2=\frac{1}{2}\int_0^{u_1}|\tilde{f}_n-\tilde{f}_{n-1}||\partial_rw_n|e^{\int_u^{u_1}2G_nds}du;\vspace{10pt}\\
&I_3=\int_0^{u_1}\frac{|J_n|}{r_n}|(h_n-\bar{h}_n)-(h_{n-1}-\bar{h}_{n-1})|e^{\int_u^{u_1}2G_nds}du;\vspace{10pt}\\
&I_4=\int_0^{u_1}\frac{|J_n-J_{n-1}|}{r_n}|h_{n-1}-\bar{h}_{n-1}|e^{\int_u^{u_1}2G_nds}du.
\end{align}
Before estimating these integrals, we'll see some useful inequalities. We have that
\[
\begin{split}
|(h_n-\bar{h}_n)+(h_{n-1}-\bar{h}_{n-1})|&\leq |h_n-\bar{h}_n|+|h_{n-1}-\bar{h}_{n-1}|\lesssim \frac{r\log^{\chi(\delta)}(e+r)}{(1+r)^{2-\delta^+}}\|w_n\|_{L_U^\infty L_r^{\infty,2-\delta}}\\
&+\frac{r\log^{\chi(\delta)}(e+r)}{(1+r)^{2-\delta^+}}\|w_{n-1}\|_{L_U^\infty L_r^{\infty,2-\delta}}\\
&\lesssim \frac{r\log^{\chi(\delta)}(e+r)}{(1+r)^{2-\delta^+}}x'.
\end{split}
\]
Also, recalling that \(\|\bar{h}\|_{L_U^\infty L_r^\infty}\leq \|h\|_{L_U^\infty L_r^\infty}\), we have
\[
|(h_n-\bar{h}_n)-(h_{n-1}-\bar{h}_{n-1})|\leq 2\|h_n-h_{n-1}\|_{L_U^\infty L_r^\infty}.
\]
The last two estimates imply that
\[
|(h_n-\bar{h}_n)^2-(h_{n-1}-\bar{h}_{n-1})^2|\leq x'\|h_n-h_{n-1}\|_{L_U^\infty L_r^\infty}\frac{r\log^{\chi(\delta)}(e+r)}{(1+r)^{2-\delta^+}}.
\]
Now, since 
\[
\frac{1}{2}\int_0^r\frac{(h_n-\bar{h}_n)^2}{s}ds\lesssim (x')^2\int_0^r\frac{s\log^{2\chi(\delta)}(e+s)}{(1+s)^{4-2\delta^+}}ds\lesssim (x')^2
\]
and similarly for the same expression with \(n-1\) instead of \(n\), then using the mean value theorem we see that for \(x'\) sufficiently small
\[
\begin{split}
|f_n-f_{n-1}|&=\left|\exp\left(\frac{1}{2}\int_0^r\frac{(h_n-\bar{h}_n)^2}{s}ds\right)-\exp\left(\frac{1}{2}\int_0^r\frac{(h_{n-1}-\bar{h}_{n-1})^2}{s}ds\right)\right|\\
&\lesssim \frac{1}{2}\int_0^r\frac{|(h_n-\bar{h}_n)^2-(h_{n-1}-\bar{h}_{n-1})^2|}{s}ds\\
&\lesssim x'\|h_n-h_{n-1}\|_{L_U^\infty L_r^\infty}\int_0^r\frac{\log^{\chi(\delta)}(e+s)}{(1+s)^{2-\delta^+}}ds\lesssim x'\|h_n-h_{n-1}\|_{L_U^\infty L_r^\infty}.
\end{split}
\]
It also follows immediately from this that
\[
\begin{split}
|\tilde{f}_n-\tilde{f}_{n-1}|&=\left|\frac{1}{r}\int_0^r(1-3s^2)(f_n-f_{n-1})ds\right|\leq \frac{1}{r}\int_0^r(1+3s^2)|f_n-f_{n-1}|ds\\
&\lesssim x'\|h_n-h_{n-1}\|_{L_U^\infty L_r^\infty}\frac{1}{r}\int_0^r(1+3s^2)ds\lesssim (1+r)^2x'\|h_n-h_{n-1}\|_{L_U^\infty L_r^\infty}.
\end{split}
\]
Now we also want a similar estimate for \(|G_n-G_{n-1}|\). We may write
\[
\begin{split}
G_n-G_{n-1}&=\frac{1}{2r}\left((f_n-f_{n-1})(1-3r^2)-\frac{1}{r}\int_0^r(f_n-f_{n-1})(1-3s^2)ds\right)\\
&=\frac{1}{2r}[f_n-f_{n-1}-\overline{f_n-f_{n-1}}]+\frac{3}{2r^2}\int_0^rs^2(f_n-f_{n-1})ds-\frac{3}{2}r(f_n-f_{n-1}).
\end{split}
\]
We have that
\[
\begin{split}
(1+r)^{2-\delta}|\partial_r(f_n-f_{n-1})|&=(1+r)^{2-\delta}\left|f_n\frac{(h_n-\bar{h}_n)^2}{2r}-f_{n-1}\frac{(h_{n-1}-\bar{h}_{n-1})^2}{2r}\right|\\
&=\frac{(1+r)^{2-\delta}}{2r}\left|(f_n-f_{n-1})(h_n-\bar{h}_n)^2+f_{n-1}((h_n-\bar{h}_n)^2-(h_{n-1}-\bar{h}_{n-1})^2)\right|\\
&\lesssim \frac{(1+r)^{2-\delta}}{2r}\left((x')^3\|h_n-h_{n-1}\|_{L_U^\infty L_r^\infty}\frac{r^2\log^{2\chi(\delta)}(e+r)}{(1+r)^{4-2\delta^+}}\right.\\
&\left.+x'\|h_n-h_{n-1}\|_{L_U^\infty L_r^\infty}\frac{r\log^{\chi(\delta)}(e+r)}{(1+r)^{2-\delta^+}}\right)\\
&\lesssim x'\|h_n-h_{n-1}\|_{L_U^\infty L_r^\infty}\log^{\chi(\delta)}(e+r)(1+r)^{\delta^+-\delta}.
\end{split}
\]
So,
\[
\begin{split}
|f_n-f_{n-1}-\overline{f_n-f_{n-1}}|&\leq \frac{1}{r}\int_0^r\int_s^r\frac{1}{(1+\rho)^{2-\delta}}(1+\rho)^{2-\delta}|\partial_r(f_n-f_{n-1})|d\rho ds\\
&\lesssim x'\|h_n-h_{n-1}\|_{L_U^\infty L_r^\infty}\frac{1}{r}\int_0^r\int_s^{r}\frac{\log^{\chi(\delta)}(e+\rho)}{(1+\rho)^{2-\delta^+}}d\rho ds\\
&\lesssim x'\|h_n-h_{n-1}\|_{L_U^\infty L_r^\infty}\frac{1}{r}\int_0^r\int_s^rd\rho ds\lesssim x'\|h_n-h_{n-1}\|_{L_U^\infty L_r^\infty}r.
\end{split}
\]
In this way we see that 
\[
\begin{split}
|G_n-G_{n-1}|&\lesssim x'\|h_n-h_{n-1}\|_{L_U^\infty L_r^\infty}+x'\|h_n-h_{n-1}\|_{L_U^\infty L_r^\infty}\frac{1}{r^2}\int_0^rs^2ds +rx'\|h_n-h_{n-1}\|_{L_U^\infty L_r^\infty}\\
&\lesssim (1+r)x'\|h_n-h_{n-1}\|_{L_U^\infty L_r^\infty}.
\end{split}
\]
Now we obtain an estimate for \(|J_n-J_{n-1}|\). We can write
\[
J_n-J_{n-1}=3(G_n-G_{n-1})+3r(f_n-f_{n-1})-\frac{1-3r^2}{4r}[(f_n-f_{n-1})(h_n-\bar{h}_n)^2+f_{n-1}[(h_n-\bar{h}_n)^2-(h_{n-1}-\bar{h}_{n-1})^2]].
\]
Therefore,
\[
\begin{split}
|J_n-J_{n-1}|&\lesssim (1+r)x'\|h_n-h_{n-1}\|_{L_U^\infty L_r^\infty}+rx'\|h_n-h_{n-1}\|_{L_U^\infty L_r^\infty}\\
&+\frac{1+3r^2}{r}\left[(x')^3\|h_n-h_{n-1}\|_{L_U^\infty L_r^\infty}\frac{r^2\log^{2\chi(\delta)}(e+r)}{(1+r)^{4-2\delta^+}}+x'\|h_n-h_{n-1}\|_{L_U^\infty L_r^\infty}\frac{r\log^{\chi(\delta)}(e+r)}{(1+r)^{2-\delta^+}}\right]\\
&\lesssim x'\|h_n-h_{n-1}\|_{L_U^\infty L_r^\infty}(1+r)\log^{\chi(\delta)}(e+r).
\end{split}
\]
Now we are ready to estimate the integrals \(I_1, I_2, I_3\) and \(I_4\). We start with \(I_1\).
\[
\begin{split}
I_1&\lesssim (x')^2\|h_n-h_{n-1}\|_{L_U^\infty L_r^\infty}\int_0^{u_1}\frac{1}{(1+r_n)^{1-\delta}}e^{\int_u^{u_1}2G_nds}du\lesssim x'\|h_n-h_{n-1}\|_{L_U^\infty L_r^\infty}\frac{1+u_1}{(1+r_1)^{2-\delta}}.
\end{split}
\]
Next we look at \(I_2\). 
\[
\begin{split}
I_2&\lesssim x'x''\|h_n-h_{n-1}\|_{L_U^\infty L_r^\infty}\int_0^{u_1}\frac{1}{(1+r_n)^{1-\delta}}e^{\int_u^{u_1}2G_nds}du\lesssim x'x''\|h_n-h_{n-1}\|_{L_U^\infty L_r^\infty}\frac{1+u_1}{(1+r_1)^{2-\delta}}.
\end{split}
\]
Now we estimate \(I_4\). 
\[
\begin{split}
I_4&\lesssim (x')^2\|h_n-h_{n-1}\|_{L_U^\infty L_r^\infty}\int_0^{u_1}\frac{\log^{2\chi(\delta)}(e+r_n)}{(1+r_n)^{1-\delta^+}}e^{\int_u^{u_1}2G_nds}du\\
&\lesssim x'\|h_n-h_{n-1}\|_{L_U^\infty L_r^\infty}\frac{\log^{2\chi(\delta)}(e+r_1)}{(1+r_1)^{2-\delta^+}}.
\end{split}
\]
Finally, we estimate \(I_3\). Using the same notation as in Proposition \ref{prop3.1}, we have
\[
\begin{split}
I_3&=\int_{\Omega_{\leq 1}}\frac{|J_n|}{r_n}|(h_n-\bar{h}_n)-(h_{n-1}-\bar{h}_{n-1})|e^{\int_u^{u_1}2G_nds}du+\int_{\Omega_{>1}}\frac{|J_n|}{r_n}|(h_n-\bar{h}_n)-(h_{n-1}-\bar{h}_{n-1})|e^{\int_u^{u_1}2G_nds}du\\
&\lesssim \|h_n-h_{n-1}\|_{L_U^\infty L_r^\infty}\int_{\Omega_{\leq 1}}\left(\frac{1+r_n}{1+r_1}\right)^{4-\eta}du\\
&+(x')^2\|h_n-h_{n-1}\|_{L_U^\infty L_r^\infty}\int_{\Omega_{>1}}\frac{[\log(e+r_n)]^{2\chi(\delta)+1-H(\delta)}}{r_n(1+r_n)^{1-2\delta^+}}e^{\int_u^{u_1}2G_nds}du\\
&\lesssim \|h_n-h_{n-1}\|_{L_U^\infty L_r^\infty}\frac{u_1}{(1+r_1)^{4-\eta}}+(x')^2\|h_n-h_{n-1}\|_{L_U^\infty L_r^\infty}\frac{[\log(e+r_1)]^{2\chi(\delta)+1-H(\delta)}}{(1+r_1)^{3-2\delta^+}}\\
&\lesssim \|h_n-h_{n-1}\|_{L_U^\infty L_r^\infty}\frac{U}{(1+r_1)^{4-\eta}}+x'\|h_n-h_{n-1}\|_{L_U^\infty L_r^\infty}\frac{[\log(e+r_1)]^{2\chi(\delta)+1-H(\delta)}}{(1+r_1)^{3-2\delta^+}}.
\end{split}
\]
Putting these estimates together we obtain
\[
\begin{split}
|w_{n+1}-w_n|(u_1,r_1)&\lesssim x'\|h_n-h_{n-1}\|_{L_U^\infty L_r^\infty}\frac{1+u_1}{(1+r_1)^{2-\delta}}+x'x''\|h_n-h_{n-1}\|_{L_U^\infty L_r^\infty}\frac{1+u_1}{(1+r_1)^{2-\delta}}\\
&+\|h_n-h_{n-1}\|_{L_U^\infty L_r^\infty}\frac{U}{(1+r_1)^{4-\eta}}+x'\|h_n-h_{n-1}\|_{L_U^\infty L_r^\infty}\frac{[\log(e+r_1)]^{2\chi(\delta)+1-H(\delta)}}{(1+r_1)^{3-2\delta^+}}\\
&+x'\|h_n-h_{n-1}\|_{L_U^\infty L_r^\infty}\frac{\log^{2\chi(\delta)}(e+r_1)}{(1+r_1)^{2-\delta^+}}\\
&\lesssim \left( \frac{U}{(1+r_1)^{4-\eta}}+\frac{[\log(e+r_1)]^{2\chi(\delta)+1-H(\delta)}}{(1+r_1)^{2-\delta^+}}(1+x'')x' \right) \|h_n-h_{n-1}\|_{L_U^\infty L_r^\infty}.
\end{split}
\]
Thus,
\[
\begin{split}
|h_{n+1}-h_n|(u,r)&\leq \int_0^r|w_{n+1}-w_n|(u,s)ds\\
&\lesssim \|h_n-h_{n-1}\|_{L_U^\infty L_r^\infty}\int_0^r\left[\frac{U}{(1+s)^{4-\eta}}+\frac{[\log(e+s)]^{2\chi(\delta)+1-H(\delta)}}{(1+s)^{2-\delta^+}}(1+x'')x'\right]ds\\
&\lesssim (U+(1+x'')x')\|h_n-h_{n-1}\|_{L_U^\infty L_r^\infty}.
\end{split}
\]
Hence, we see that if \(\|h_0'\|_{L_U^\infty L_r^{\infty,2}}\) and \(U\) are small enough we can get the implicit constant smaller that \(1\), thereby showing that the sequence \((h_n)_n\) contracts in \(L_U^\infty L_r^\infty\). 
\end{proof}
\end{document}